\documentclass[prd, twocolumn, nofootinbib]{revtex4}
\usepackage{graphicx, epsfig}

\newcommand{\mpvectorfig}{Tegmark_2-8.color.ps}

\newcommand{\website}{{\tt http://www.phys.cwru.edu/projects/mpvectors/}}

\def\ntwo{{\bf\hat n}_{2}}
\def\nthree{{\bf\hat n}_{3}}
\def\nl{{\bf\hat n}_{\ell}}
\def\alm{a_{\ell m}}

\begin{document}

\title{Multipole Vectors---a new representation of the CMB sky \\
and evidence for statistical anisotropy or non-Gaussianity at $2\leq\ell\leq8$}

\author{Craig J. Copi}
\affiliation{Department of Physics,
Case Western Reserve University,
Cleveland, OH~~44106-7079}
\homepage[More information, color pictures, and code are
available at ]{http://www.phys.cwru.edu/projects/mpvectors/}

\author{Dragan Huterer}
\affiliation{Department of Physics,
Case Western Reserve University,
Cleveland, OH~~44106-7079}

\author{Glenn D. Starkman}
\affiliation{Theory Division CERN, 
Geneva, Switzerland}
\affiliation{Department of Physics,
Case Western Reserve University,
Cleveland, OH~~44106-7079}

\begin{abstract}
  We propose a novel representation of cosmic microwave anisotropy maps,
  where each multipole order $\ell$ is represented by $\ell$ unit vectors
  pointing in directions on the sky and an overall magnitude.  These
  ``multipole vectors and scalars" transform as vectors under rotations.
  Like the usual spherical harmonics, multipole vectors form an irreducible
  representation of the proper rotation group $SO(3)$.  However, they are
  related to the familiar spherical harmonic coefficients, $\alm$, in a
  nonlinear way, and are therefore sensitive to different aspects of the
  CMB anisotropy.  Nevertheless, it is straightforward to determine the
  multipole vectors for a given CMB map and we present an algorithm to
  compute them.  A code implementing this algorithm is available at
  \website.  Using the WMAP full-sky maps, we perform several tests of the
  hypothesis that the CMB anisotropy is statistically isotropic and
  Gaussian random. We find that the result from comparing the oriented area
  of planes defined by these vectors between multipole pairs $2\leq \ell_1
  \ne \ell_2\leq 8$ is inconsistent with the isotropic Gaussian hypothesis
  at the 99.4\% level for the ILC map and at 98.9\% level for the cleaned
  map of Tegmark {\it et al.}  A particular correlation is suggested
  between the $\ell=3$ and $\ell=8$ multipoles, as well as several other
  pairs.  This effect is entirely different from the now familiar planarity
  and alignment of the quadrupole and octupole: while the aforementioned is
  fairly unlikely, the multipole vectors indicate correlations not expected
  in Gaussian random skies that make them unusually likely.  The result
  persists after accounting for pixel noise and after assuming a residual
  10\% dust contamination in the cleaned WMAP map.  While the definitive
  analysis of these results will require more work, we hope that multipole
  vectors will become a valuable tool for various cosmological tests, in
  particular those of cosmic isotropy.
\end{abstract}

\maketitle

\section{Introduction}

A great deal of attention is currently being devoted to examining the
power spectrum of the cosmic microwave background (CMB) temperature
anisotropies $\Delta T(\Omega)/T$ extracted from the
Wilkinson Microwave Anisotropy Probe
(WMAP)~\cite{wmap_results,wmap_foreground,wmap_angps,wmap_low} and other
CMB data~\cite{maxima,DASI,acbar,CBI,vsa,archeops,boom}.
Decomposing the temperature in spherical harmonics
\begin{equation}
  \frac{\Delta T (\Omega)}{T} = \sum_{\ell m} \alm Y_{\ell m} (\Omega)
\end{equation}
and deducing the angular power spectrum
\begin{equation}
  C_\ell \equiv \frac{1}{2\ell+1} \sum_{m=-\ell}^{\ell} \left\vert
    \alm\right\vert^2,
\end{equation}
as a function of $\ell$ allows cosmologists to fit the parameters of
cosmological models to unprecedented accuracy, possibly even probing
the physics of the inflationary epoch.  Similarly, the power spectrum
of the temperature-polarization cross-correlation function is teaching
us about the physics of the reionization of the universe, presumably
by the first generation of stars; and the power spectrum of the
temperature-galaxy cross-correlation function is teaching us about the
statistical distribution of matter in the universe.  But does the only
physics lie in the angular power spectra?  Is the sky statistically
isotropic, so that any variation in the values of the individual
multipole moments $\alm$ with fixed $\ell$ represents only statistical
fluctuations, or could there be subtle correlations between the
$\alm$?  If the sky is statistically isotropic, is it Gaussian---are
the $\alm$ of each fixed $\ell$ drawn from a Gaussian distribution of
variance that is only $\ell$-dependent?  Are there other interesting
deviations from the simplest picture?

In the standard inflationary cosmology the answer to the question just
posed is that in the linear regime, {\it i.e.}\ at low $\ell$, the $\alm$
are realizations of Gaussian random variables of zero mean, with variances
that depend only on $\ell$ (statistical isotropy).  This paradigm is so
strongly believed, both because of considerable observational evidence and
considerable theoretical prejudice, that relatively little (though some,
e.g.~\cite{scott}) attention has been paid to searches for deviations from
statistical isotropy.

In this paper we set out to search for one particular deviation---special
directions on the sky.  We do this by first constructing from each
multipole moment
\begin{equation}
  \frac{ \Delta T_\ell(\Omega)}{T} = \sum_{m=-\ell}^{\ell}
  \alm Y_{\ell m}(\Omega)
\end{equation}
of the CMB sky a set of $\ell$ unit vectors $\{\hat v^{(\ell,i)} \mid
i=1,...,\ell\}$ and a scalar $A^{(\ell)}$ that completely characterize that
multipole. We then examine the correlations between pairs of such sets of
vectors, $\{\hat v^{(\ell_1,i_1)}\}$ and $\{\hat v^{(\ell_2,i_2)}\}$,
comparing them with Monte Carlo simulations of CMB skies with statistically
isotropic Gaussian random $\alm$.  If the sky is statistically isotropic
with Gaussian random $\alm$, $\left\langle \alm a_{\ell'm'}\right\rangle =
C_\ell \delta_{\ell\ell'} \delta_{m m'}$.  Since the $\{\hat v^{(\ell_i,i_i)}\}$
depend only on the $a_{\ell_im}$, $\{\hat v^{(\ell_1,i_1)}\}$, and $\{\hat
v^{(\ell_2,i_2)}\}$ should be uncorrelated for $\ell_1\neq\ell_2$.

We constructed these vectors for $2\leq \ell \leq 8$ for a set of full-sky
maps including the WMAP Internal Linear
Combination~(ILC)~\cite{wmap_results}, and the WMAP map as cleaned by
Tegmark {\it et al.}~\cite{tegmark_cleaned}.  We applied four statistical
tests to the set of vectors from these full-sky maps.  We find that one of
the tests is inconsistent with the hypothesis of statistical isotropy and
Gaussianity at the 99\% confidence level.  This work complements recent
work by Eriksen {\it et al.}~\cite{NorthSouth} looking at north-south
asymmetries in N-point functions, work by Park \cite{park} looking at genus
curves, work by Hajian {\it et al.}~\cite{Hajianetal} using the
so-called $\kappa_\ell$ test \cite{HajianSouradeep} which finds violations
of statistical isotropy for $15\leq\ell\leq45$, and work by Vielva {\it et
  al.}~\cite{vielva} using the spherical mexican hat wavelet technique
where they found a strong signal for non-Gaussianity.

\section{Motivation for a New Test of  Statistical Isotropy and Gaussianity}

Tests of non-Gaussianity, as opposed to statistical isotropy, have a
long and rich history.  Motivation for those tests originally came
from the realization that non-Gaussianity is a signature of structure
formation by topological defects~\cite{defects}, while inflation
predicts Gaussian CMB anisotropies.  Subsequently it has been realized
that, even if inflation seeded the structure in the universe, CMB
non-Gaussianity may be present as a signature of features in the
inflationary model~\cite{salopek_bond,falk,gangui,wang_kam,bern_uzan}.
Finally, late-time processes in the universe will induce
non-Gaussianity on small scales
\cite{luo_schramm,luo,munshi,gold_spergel,cooray_hu,cooray_sz}. The
tests of non-Gaussianity include studies of the bispectrum and
skewness~\cite{heavens,spergel_gold,verde,komatsu_cobe,
santos_prl,santos_big,komatsu_wmap},
trispectrum~\cite{hu_trispec,detroia}, Minkowski functionals and the
genus statistic~\cite{gott,smoot_genus,winitzki,dolgov,park_teg,park},
spherical wavelets~\cite{barreiro,mukherjee,vielva}, a combination of
these~\cite{kogut,bromley,wu,savage,polenta}, and many other
methods~\cite{hansen,dore,chiang,coles}.  Of particular interest was
the claim for non-Gaussianity in the COBE 4-year data~\cite{ferreira},
but this was shown to be an artifact of a particular known
systematic~\cite{zaroubi}.  Nevertheless, efforts to test the
Gaussianity of the CMB continue, and most, though not all
(e.g.~\cite{magueijo,park}), have so far given results entirely in
agreement with the Gaussian hypothesis.

As these previous studies have shown, it is a challenge to test such
fundamental assumptions as statistical isotropy and Gaussianity
without theoretical direction on what deviations to expect---they can
be violated in a very large number of ways each of which could easily
be hidden from the test that was actually performed.  An instructive
example is the effects on the CMB of any non-trivial topology of the
universe, for example a universe which is a 3-torus.  If the length
scale of cosmic topology (for example the length of the smallest
non-trivial closed curve) is sufficiently short, this will manifest
itself in various two-point temperature-temperature correlations, such
as the so-called ``circles-in-the-sky" signature \cite{CSS}: if there
are closed paths shorter than the diameter of the last scattering
surface, then the last scattering surface will self-intersect along
circles.  These circles can be viewed by an observer from both
sides---from one side in one direction on the sky, and from the other
side in some other direction.  The temperature as a function of
location around the circle as seen from the two sides will be very
strongly correlated.  One can therefore search for such pairs of
circles.  A definitive direct search is currently being conducted by
the original proponents of the signature.  However, for us this
example serves to show precisely why it is so difficult to perform a
comprehensive test of statistical isotropy and Gaussianity.  The
topology-induced temperature correlations are strong only on or very
near the matched circle pairs, thus the multipole coefficients $\alm$,
and other statistics that are weighted averages over the entire sky,
would be poor tools for searching for such circles.  Thus testing
these phenomena is in part a matter of continually searching for
(preferably) physically motivated ways in which they manifest.

The CMB data itself may provide motivation for searching for deviations from
the standard inflationary predictions (of statistical isotropy and
Gaussianity), especially at large angular scales.  An absence of
large angular scale correlations in the CMB sky relative to the
inflationary prediction was first noted by COBE DMR in their first year
data \cite{DMR1}, which showed what was reported as an anomalously low
quadrupole, $C_2$.  Because the cosmic variance in the quadrupole is quite
large, it was widely dismissed as a statistical anomaly.  The result
persisted and was strengthened by the COBE DMR four year data~\cite{DMR4}.
The recent WMAP analysis shows a marked absence of power on scales
extending from $60^\circ$ to $180^\circ$ to an extent that cannot be
explained solely by a low quadrupole~\cite{wmap_low}.  Note that the
estimator used by the WMAP team has been shown to be non-optimal when
applied to incomplete maps of the sky.  When an alternative estimator is
applied~\cite{tegmark_cleaned,efstathiou} or a full-sky map is
analyzed~\cite{wmap_results} the discrepancy becomes less significant.

An absence of power on large scales is expected in some topologically
non-trivial universes.  In a compact universe there is a spectral cutoff of
long wavelength modes leading to a suppression of power near this cutoff.
One method of looking for such a cutoff is the ``circles-in-the-sky''
signature as noted above.  In general, such compact topologies would lead to
``special directions'' in the universe.  To search for special directions we
need a method of defining our directions.  Our definition, as discussed in
the next section, is to decompose the $\ell^{\rm th}$ multipole into $\ell$
unit vectors, these vectors are then studied to search for peculiar
alignments.  These vectors contain the full information of the $\alm$ but
encode it in a different way that allows for one to more easily look for
special directions.  In particular, the components of these vectors are
non-linear combinations of the $\alm$ (for a fixed $\ell$).

\section{Defining ``Special Directions"}

One attempt to look at the statistical isotropy of the CMB on large scales
was the analysis of the quadrupole and octupole moments of the WMAP sky by
de Oliveira-Costa {\it et al.}~\cite{angelica}.  They found that the
quadrupole was unusually small, and that the octupole was unusually planar
and unusually aligned with the quadrupole.  They identified an axis with
each multipole by finding, for each $\ell$, the axis $\nl$ around which the
angular momentum dispersion
\begin{equation}
\left\langle {\Delta T\over T}(\nl) 
     \left\vert \,\left(\nl \cdot  {\bf \rm L}\right)^2  \,\right\vert
{\Delta T\over T}(\nl) \right\rangle = 
\sum_{m=-\ell}^{\ell} m^2 \vert \alm (\nl)\vert^2 
\end{equation}
is maximized.  (Here $\alm (\nl)$ are the spherical
harmonic coefficients of the CMB map in a coordinate system with its
z-axis in the $\nl$ direction.)  They found that the $\ntwo$ and the
$\nthree$ directions:
\begin{eqnarray}
\ntwo = (-0.1145, -0.5265, 0.8424)&  \quad {\rm and} \nonumber \\
\nthree = (-0.2578, -0.4207, 0.8698)& 
\end{eqnarray}
are unusually aligned---their dot product is $0.9838$.  This has only a $1$
in $62$ chance of happening if $\ntwo$ and $\nthree$ are uncorrelated and
the dot product is uniformly distributed on the sky.  De Oliveira-Costa
{\it et al.}~\cite{angelica} point out that these values of $\ntwo$ and
$\nthree$ could be explained by a universe which has a compact direction
parallel to $\ntwo$ and $\nthree$ and of length approximately equal to the
horizon radius; but this is ruled out by other tests, including the absence
of matched circles in these directions.

It has been noticed that the quadrupole and the octupole in the
cleaned WMAP skies remain dominated by a hot and a cold spot in the
Galactic plane---one in the general direction of the Galactic center,
and the other in the general direction of the molecular cloud in
Taurus.  This raises the possibility that the observed correlation is
dominated by foreground contamination.  One would like therefore to
examine in more detail the correlations between the $\alm$
corresponding to possible preferred directions, or correlations in
directions between the various multipoles.

The question then is how best to associate directions with the CMB
multipoles.  De Oliveira-Costa {\it et al.}~\cite{angelica} associated only
one direction with each multipole, corresponding to two real degrees of
freedom, whereas the $\alm$ of a given $\ell$ have $2\ell+1$ real degrees of
freedom.  The $\ell^{\rm th}$ multipole, $f_\ell(\Omega)$, in the multipole
expansion of a function $f(\Omega)$ on a sphere
\begin{equation}
f (\Omega) \equiv \sum_{\ell} f_{\ell}(\Omega) \equiv  
\sum_{\ell m} \alm Y_{\ell m} (\Omega)
\end{equation}
can be fully represented by a symmetric, traceless rank $\ell$ tensor,
$F_{i_1 ... i_\ell}$ ($i_k = 1,2,3$).  Such a tensor can readily be
constructed from the outer product of $\ell$ unit vectors,
$\hat v^{(\ell,i)}$, and a single scalar, $A^{(\ell)}$.  (Strictly
speaking these are headless vectors, i.e.\ points on the projective
two-sphere.  The sign of each vector can always be absorbed by the scalar.
The sign of the scalar takes on physical signficance when we define a
convention for the multipole vectors, such as, all of them point in the
northern hemisphere.)

\subsection{Vector Decomposition}
\label{sec:vectordecomp}

The correspondence between these vectors and the usual multipole
coefficients can readily be seen for a dipole.  A dipole defines a
direction in space---the line along which the dipole lies.  The
standard correspondence is
\begin{equation}
  Y_{1,0} \rightarrow \hat z,\quad Y_{1,\pm1} \rightarrow \mp
  \frac{1}{\sqrt 2} \left ( \hat x \pm i\hat y \right).
 \label{eqn:dipoledir}
\end{equation}
Thus
\begin{eqnarray}
  &&\sum_{m=-1}^1 a_{1,m}Y_{1,m}(\Omega) \nonumber\\[0.15cm] 
 &=& A^{(1)}(\hat v^{(1,1)}_x , \hat v^{(1,1)}_y, \hat v^{(1,1)}_z)
\cdot (\sin\theta \cos\phi, \sin\theta \sin\phi, \cos\theta)\nonumber \\[0.2cm]
   &\equiv& A^{(1)}\hat v^{(1,1)} \cdot \hat e,
\end{eqnarray}
where $\hat e$ is the radial unit vector in spherical coordinates.
For a real valued function, the vector's components are found to be
\begin{equation}
  v^{(1,1)}_x = -\sqrt{2} a_{1,1}^{\rm re}, \quad 
  v^{(1,1)}_y = \sqrt{2} a_{1,1}^{\rm
    im}, \quad v^{(1,1)}_z = a_{1,0},
\end{equation}
and $A^{(1)} = |\vec v^{(1,1)}|$ (which can then be used to construct the
unit vector, $\hat v^{(1,1)}$).

To extend this to the $\ell^{\rm th}$ multipole, we want to write
heuristically that
\begin{equation}
\label{eqn:heuristic}
\sum_{m=-\ell}^{\ell} \alm Y_{\ell m}(\Omega) \approx A^{(\ell)} \left
( \hat v^{(\ell,1)}\cdot \hat e \right) \cdots \left ( \hat
v^{(\ell,\ell)}\cdot \hat e \right),
\end{equation}
for each of the $\ell$ directions given by $\hat e$.  This cannot be quite
right since the product of $\ell$ vectors would contain not only components
of angular momentum $\ell$, but also of angular momenta $\ell-2$, $\ell-4$,
{\it etc}.  However, a simple power-counting shows that, once the reality
conditions have been imposed on $a_{\ell m}$, $\ell$ unit vectors and a
scalar contain the same number of degrees of freedom as does $a_{\ell m}$,
namely, $(2\ell+1$) real degrees of freedom.  We therefore expect that the
components of lower angular momentum found in the right hand side of
equation (\ref{eqn:heuristic}) are not independent.  We shall see this
explicitly in equations (\ref{eqn:peel}) and (\ref{eqn:system}) below, and
more elegantly in section \ref{secn:elegant}.  For now, let us treat
equation (\ref{eqn:heuristic}) as motivation and proceed.

Instead of solving (\ref{eqn:heuristic}) directly for all $\hat
v^{(\ell,i)}$ we peel off one vector at a time, finding first a vector
$\hat v^{(\ell,1)}$ (with components $\hat v^{(\ell,1)}_{i_1}$), and a
rank $\ell-1$ symmetric, traceless tensor, $a^{(\ell,1)}$.  We can
think of $i_1$ as running over $x$, $y$, and $z$, or more conveniently
over $-1$, $0$, and $1$.  Similarly, we can write $a^{(\ell,1)}$ as a
$3\times3\times\cdots\times3$ ($(\ell-1)$ terms) matrix,
$a^{(\ell,1)}_{i_2\cdots i_\ell}$; however, this hides its traceless,
symmetric nature and makes it appear that $a^{(\ell,1)}$ has far more
independent degrees of freedom than it actually does.  It is therefore
more instructive to write the $2\ell-1$ independent components as
$a^{(\ell,1)}_{\ell-1,m}$, with $m=-(\ell-1),...,(\ell-1)$.

We repeat this procedure recursively on the remaining symmetric, traceless
tensor from the previous step.  Thus we next peel from $a^{(\ell,1)}$ a
vector, $\hat v^{(\ell,2)}$, and a
rank $\ell-2$ symmetric, traceless tensor $a^{(\ell,2)}$, and repeat
until we have found the full set of $\ell$ vectors, $\{\hat
v^{(\ell,i)} \mid i=1,...,\ell\}$.  The scalar $A^{(\ell)}$, is found in
the last step when the second to last vector $\hat v^{(\ell,\ell-1)}$ is
peeled off.  In this case the remaining symmetric, traceless tensor,
$a^{(\ell,\ell-1)}$ is rank $1$, and is just the product of the final
unit vector $\hat v^{(\ell,\ell)}$ and the scalar $A^{(\ell)}$.

To apply the recursive procedure outlined above we use the following
recursion relation to peel off one vector
\begin{equation}
  Y_{1,j} Y_{\ell-1,m-j} = C_j^{(\ell,m)} Y_{\ell m} + D_j^{(\ell,m)}
  Y_{\ell-2,m} \label{eqn:recursion}
\end{equation}
for $j=-1, 0, 1$ where
\begin{eqnarray}
  C_0^{(\ell,m)} &=& \sqrt{\frac{3}{4\pi}}\sqrt{\frac{(\ell-m) (\ell+m)}{(2l-1)
      (2l+1)}}  \nonumber \\
  C_{\pm1}^{(\ell,m)} &=& \sqrt{\frac{3}{8\pi}}\sqrt{\frac{(\ell\pm m-1) 
      (\ell\pm
      m)}{(2l-1) (2l+1)}}  \nonumber \\
  D_0^{(\ell,m)} &=&
  \sqrt{\frac{3}{4\pi}}\sqrt{\frac{(\ell-m-1) (\ell+m-1)}{(2l-3) (2l-1)}}
  \\
  D_{\pm1}^{(\ell,m)} &=& \nonumber
  -\sqrt{\frac{3}{8\pi}}\sqrt{\frac{(\ell\mp m-1) (\ell\mp m)}{(2l-3) (2l-1)}}.
\label{eqn:Clebsch}
\end{eqnarray}
For a given $\ell$ we peel off a vector using
\begin{eqnarray}
  \sum_{m=-\ell}^\ell \alm Y_{\ell m}
   &=& \sum_{\tilde m=-(\ell-1)}^{(\ell-1)} 
  \sum_{j=-1}^1 {\hat v}^{(\ell,1)}_{j}
  a^{(\ell,1)}_{\ell-1,\tilde m} Y_{1,{j}}Y_{\ell-1,\tilde m}  \nonumber\\
  &-& \sum_{m'=-(\ell-2)}^{\ell-2} b^{(\ell,1)}_{m'}
  Y_{\ell-2,m'}
  \label{eqn:peel}
\end{eqnarray}
and that
\begin{equation}
  \left| {\hat v}^{(\ell,1)}\right| = 1 .
\end{equation}
The second term on the right-hand side of equation (\ref{eqn:peel}), is
required to guarantee that this rank $\ell-1$ tensor is traceless.  In
other words, it subtracts off the trace.  We can see this is necessary from
the $Y_{\ell-2,m}$ term in the recursion relation~(\ref{eqn:recursion}).
The presence of this $\ell-2$ term is as anticipated in the discussion
following equation (\ref{eqn:heuristic})

Plugging in the recursion relation~(\ref{eqn:recursion}) yields
$4\ell-1$ coupled quadratic equations that must be solved for the
$4\ell-1$ unknowns: $\hat v^{(\ell,1)}_j$, $a^{(\ell,1)}_{\ell-1,m-j}$, and
$b^{(\ell,1)}_{m'}$.  From them we find
\begin{eqnarray}
\label{eqn:system}
  \alm = \sum_{j=-1}^1 C_j^{(\ell,m)} \hat v_j^{(\ell)} 
  a_{\ell-1,m-j}^{(\ell, 1)} &\quad&
  \hbox{($2l+1$ equations)}, \nonumber \\
  b^{(\ell,1)}_{m'} = \sum_{j=-1}^1 D_j^{(\ell,m')} \hat v_j^{(\ell)}
  a_{\ell-1,m'-j}^{(\ell, 1)} &\quad&
  \hbox{($2l-3$)}, \label{eqn:vectordecomp}\\
  \left| \hat v^{(\ell)} \right| = 1 &&
                \hbox{(1 equation)}. \nonumber
\end{eqnarray}
These equations are easily solved numerically, though see
appendix~\ref{app:vectordecomp}.  

Notice that, again as anticipated  below equation (\ref{eqn:heuristic}),
the components, $b^{(\ell,1)}_{m'}$ are
not relevant for further calculations; they are functions of the $\hat
v^{(\ell,1)}_{i_1}$ and $a^{(\ell,1)}_{\ell-1,m-{i_1}}$ and are not
independent.  This means that the vectors $v^{\ell,i}$ we calculate
are indeed unique as claimed.

\begin{figure*}
\includegraphics[width=7in]{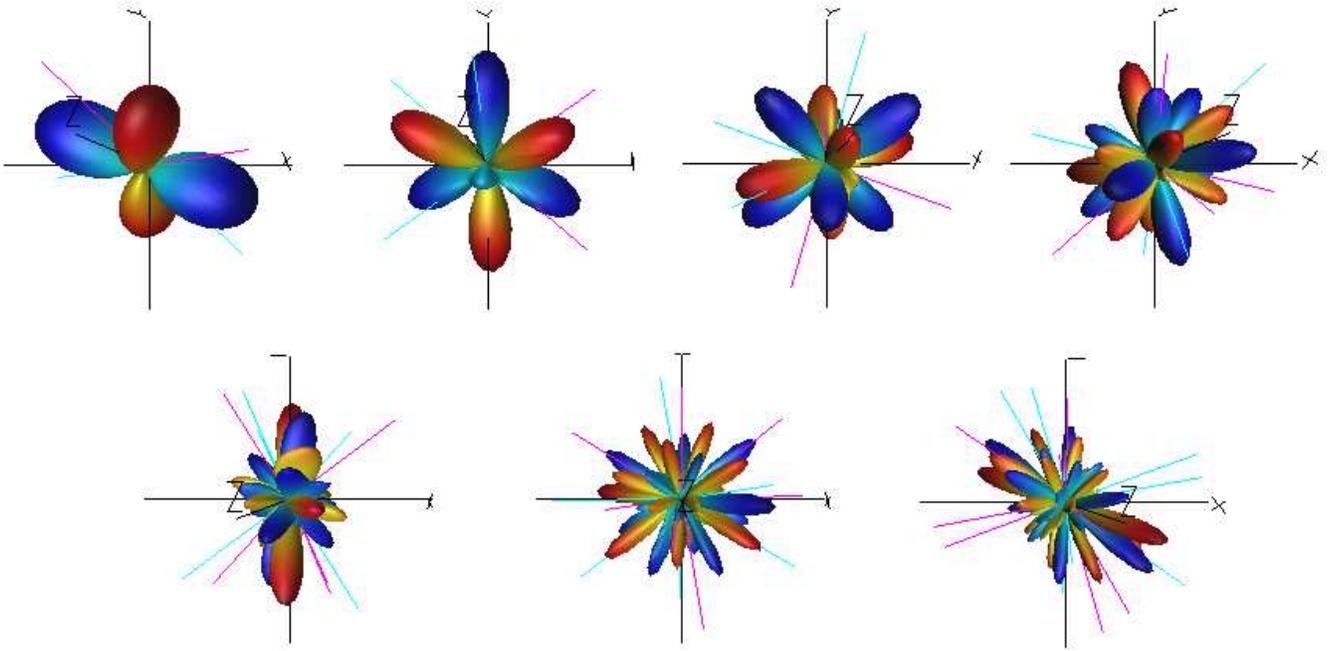}
\caption{
  An image of the sky as decomposed into the $\ell=2$ through $8$ multipole
  moments based on the first year WMAP results~\cite{wmap_results} as
  cleaned by Tegmark {\it et al.}~\cite{tegmark_cleaned}.  Shown are the
  $\sum_m \alm Y_{\ell m}$ and the vectors calculated for these multipoles.
  The vectors are drawn as ``sticks'' since they only defined up to a
  sign (thus they are ``headless'' vectors).  See \website\ for a
  full sized color picture.}
\label{fig:mpvectors}
\end{figure*}

The general procedure now follows: from the given $\alm$ we construct
$\hat v^{(\ell,1)}$ and $a^{(\ell, 1)}_{\ell-1,m}$.  We continue
using $a^{(\ell, 1)}_{\ell-1,m}$ to find $\hat v^{(\ell,2)}$ and
$a^{(\ell,2)}_{\ell-2,m}$.  This is repeated until we find
$a^{(\ell, \ell-2)}_{2,m}$ which gives the final two vectors $\hat
v^{(\ell,\ell-1)}$ and $\hat v^{(\ell,\ell)}$.  The result of this
-procedure for the CMB is shown in figure~\ref{fig:mpvectors}.

\subsection{Alternative derivation of vectors} 
\label{secn:elegant}

In practice we have implemented the procedure described above
(appendix~\ref{app:vectordecomp}) and solved the set of
equations~(\ref{eqn:vectordecomp}) for the analysis we have performed.
A mathematically more sophisticated decomposition procedure that leads to
the same set of vectors without the need to calculate the $b_{m'}$ begins
by recognizing that  
\begin{eqnarray}
 f_\ell(\Omega) &=& F^{(\ell)}_{i_1 \cdots i_\ell}  \left[ \hat e^{i_1} 
   \hat e^{i_2} \cdots \hat e^{i_\ell}\right] 
 \equiv F^{(\ell)}_{i_1 \cdots i_\ell} {\cal O}^{i_1\cdots i_\ell}
 \\[0.2cm]
&=& A^{(\ell)}\left [\hat v^{(\ell,1)}_{i_1} \hat v^{(\ell,2)}_{i_2}  \cdots 
    \hat v^{(\ell,\ell)}_{i_\ell}\right]
  \left[ \hat e^{i_1} \hat e^{i_2} \cdots \hat e^{i_\ell}\right]. \nonumber
\end{eqnarray}
Here $\hat e$ is again a radial unit vector, there is an implicit sum over
repeated indices ($i_1, \cdots, i_\ell$), and the square brackets represent
the symmetric trace free part of the outer product.  For a general
symmetric tensor, $S^{(i_1\cdots i_\ell)}=S^{i_1\cdots i_\ell}$,
\begin{eqnarray}
\left [S^{i_1 ... i_\ell}\right]  &\equiv& 
 \frac{(\ell!)^2}{(2 \ell)!} \sum_{k=0}^{\lfloor \ell/2 \rfloor}  (-1)^k 
 \frac{(2\ell-2k)!}{ k! (\ell - k)! (\ell - 2 k)!} \nonumber\\[0.1cm]
 &\times&\delta^{(i_1 i_2
 } \cdots \delta^{i_{2k-1} i_{2k} 
 } 
 S^{
   i_{2k+1} \cdots i_\ell) p_1 \cdots p_{2k} } \\[0.1cm]
&\times&\delta_{p_1 p_2} \cdots \delta_{p_{2k-1} p_{2k}}, \nonumber
\end{eqnarray}
where there is an implicit sum over the repeated indices $p_1,...,p_{2k}$,
and  $()$ denotes symmetrization  of the enclosed indices :
\begin{equation}
T^{(i_1\cdots i_\ell) }
\equiv \frac{1}{\ell!} \sum_{\pi \in U_\ell} T^{i_{\pi(1)} \cdots i_{\pi(\ell)}
} .
\end{equation}
($U_\ell$ is the group of permutations of the numbers $1,...,\ell$.)
The particular combination 
\begin{equation}  
{\cal O}^{i_1\cdots i_\ell} \equiv \left[ \hat e^{i_1} \hat e^{i_2} \cdots
  \hat e^{i_\ell}\right] 
\end{equation}
simplifies considerably because $\hat e^a \hat e^b\delta_{ab} =1$
\begin{eqnarray} 
{\cal O}^{i_1\cdots i_\ell}  
&=& \frac{(\ell!)^2}{(2 \ell)!} \sum_{k=0}^{\lfloor\ell/2\rfloor} \left[ (-1)^k 
\frac{(2\ell-2k)!}{ k! (\ell - k)! (\ell - 2 k)!} \right . \nonumber \\
&&\times \left . \delta^{(i_1 i_2
  } \cdots \delta^{i_{2k-1} i_{2k}
  } \hat e^{i_{2k+1}} \cdots \hat e^{
    i_\ell)} \right].
\end{eqnarray}
More importantly, it is easily calculable because of the
recursion relation:
\begin{equation} \left[ \hat e^{i_1} \cdots \hat e^{i_{(j+1)}} \right] = 
  \left[ \hat
  e^{i_1} \left[ \hat e^{i_2} \cdots \hat e^{i_{(j+1)}} \right]
\right ] \end{equation}
Similarly, the individual $\hat v^{(\ell,i)}$, can be peeled off one by one
by recursion:
\begin{eqnarray}
F^{(\ell)}_{i_1 \cdots i_\ell}\nonumber &=&   
\left[ \hat v^{(\ell,1)}_{i_1}, a^{(\ell,1)}_{i_2 \cdots i_\ell} \right]\\ &=& 
\left[ \hat v^{(\ell,1)}_{i_1},\left[ \hat v^{(\ell,2)}_{i_2},
    a^{(\ell,2)}_{i_3 \cdots i_\ell} \right]\right] \\
& = &A^{(\ell)}\left [\hat v^{(\ell,1)}_{i_1}  
  \hat v^{(\ell,2)}_{i_2}  \cdots 
  \hat v^{(\ell,\ell)}_{i_\ell}\right]\nonumber
\label{eqn:newpeel}
\end{eqnarray}
assuming that the $F^{(\ell)}_{i_1 \cdots i_\ell} $ can be
calculated.  But, these are easily represented as integrals over the
sky (just as are the $\alm$):
\begin{equation}
F^{(\ell)}_{i_1 \cdots i_\ell} = 
\frac{(2\ell + 1) (2\ell)! }{(4\pi) 2^\ell (\ell!)^2} \int_{S_2} 
\frac{\Delta T(\Omega)}{T} {\cal O}^{i_1\cdots i_\ell}
d\Omega
\end{equation}
Alternatively, there is an explicit (analytic) relation between the
$Y_{\ell m}$ and the $O^{i_1\cdots i_\ell}$, so that $F^{(\ell)}_{i_1
\cdots i_\ell} $ can be expressed in terms of $\alm$.  Thus, given
$F^{(\ell)}$ calculated from the sky, equation (\ref{eqn:newpeel}) is
a sequence of $2\ell+1$ coupled quadratic equations for the
$\hat v^{(\ell,i)}$ (and $A^{(\ell)}$) analogous to equation
(\ref{eqn:peel}), in which the $b^{(\ell,i)}$ have been eliminated.

\section{Sources of Error and Accuracy in Determining the Multipole Vectors}
\label{sec:error}

With the actual full-sky CMB maps, such as those that we use, the main
source of error is pixel noise which accounts for the imperfections in
measured temperature on the sky. Pixel noise depends on a variety of
factors, one of which is the number of times a given patch has been
observed. Pixel noise for WMAP is reported as $\sigma_{\rm
pix}=\sigma_0/\sqrt{N_{\rm obs}}$, where $\sigma_0$ is noise per
observation and $N_{\rm obs}$ is the number of observations per given
pixel~\cite{wmap_suppl}.  For the WMAP V-band map the reported noise
per pixel is $\sigma_0=3.11\;{\rm mK}$ and the variation in the number
of observations of each pixel is moderately small. Nevertheless,
it is important to account for the inhomogeneous distribution of pixel
noise, as we do in our full analysis in Secs.~\ref{sec:tests} and
\ref{sec:results} (where we find that the inhomogeneity of the pixel
noise does not significantly change the main results).  For the
purposes of illustrating the accuracy in determining the multipole
vectors, however, we assume a homogeneous noise with the measured mean
value of $N_{\rm obs}= 490$ observations per pixel.

The calculation of the pixel noise is straightforward. For equal-area
pixels (as in Healpix~\cite{healpix}) subtending a solid angle
$\hat\Omega_{\rm pix}$
and assuming a full-sky map we have
\begin{eqnarray}
\alm&=&\int d\Omega \,Y_{\ell m}^*(\Omega) \,\Delta T(\Omega)\nonumber\\[0.2cm]
&=&\hat\Omega_{\rm pix}\sum_{\hat{\Omega}} 
       Y_{\ell m}^*(\hat{\Omega})\,\Delta T(\hat{\Omega})
\end{eqnarray}
so that, using $\sigma_{\rm pix}^2\equiv \left\langle\left(\Delta
T(\hat\Omega)\right)^2 \right\rangle\approx {\rm constant}$, we have
\begin{eqnarray}
&&\langle\alm a_{\ell' m'}^*\rangle \nonumber \\[0.2cm]
&=&  \sum_{\hat{\Omega}} \hat\Omega_{\rm pix} Y_{\ell m}^*(\hat{\Omega})
     \sum_{\hat{\Omega}'} \hat\Omega_{\rm pix} Y_{\ell' m'}(\hat{\Omega}')
   \left\langle\Delta T(\hat{\Omega})\Delta T^*(\hat{\Omega}')
   \right\rangle \nonumber\\[0.1cm]
&=& \hat\Omega_{\rm pix}\sum_{\hat{\Omega}} \hat\Omega_{\rm pix}
   Y_{\ell m}^*(\hat{\Omega})
    Y_{\ell m}^*(\hat{\Omega})\nonumber
    \left\langle\left(\Delta T(\hat{\Omega})\right)^2\right\rangle 
    \,\delta_{\ell\ell'}\delta_{mm'} \nonumber\\[0.1cm]
& \simeq & \hat\Omega_{\rm pix}\sigma_{\rm pix}^2 \,\delta_{\ell\ell'}
    \delta_{mm'}.
\end{eqnarray}

Using the V-band map parameters, the pixel noise for {\tt NSIDE=512}
Healpix resolution is $\sqrt{\hat\Omega_{\rm pix}}\,\sigma_{\rm
  pix}=2.7\times 10^{-4}$ mK.  We adopt this quantity as an estimate of
pixel noise for {\it all\/} maps we use.

\subsection{Accuracy in Determining the Multipole Vectors}

We would like to find how accurately the multipole vectors are
determined. To do this, we add a Gaussian-distributed noise with
standard deviation $\sigma_{\alm}$ to each $\alm$ 
\begin{equation}
\alm \rightarrow \alm+\mathcal{N}(0, \sigma^2_{\alm})
\label{eq:pixnoise}
\end{equation}
where $\mathcal{N}(\mu, \sigma^2)$ denotes a Gaussian random variate
with mean $\mu$ and variance $\sigma^2$.  We repeat this many times in order to
determine the distribution in the noise-added multipole vectors and 
their spherical coordinates $(\theta,\phi)$.

\begin{figure}
\includegraphics[height=3.5in, width=2.5in, angle=-90]{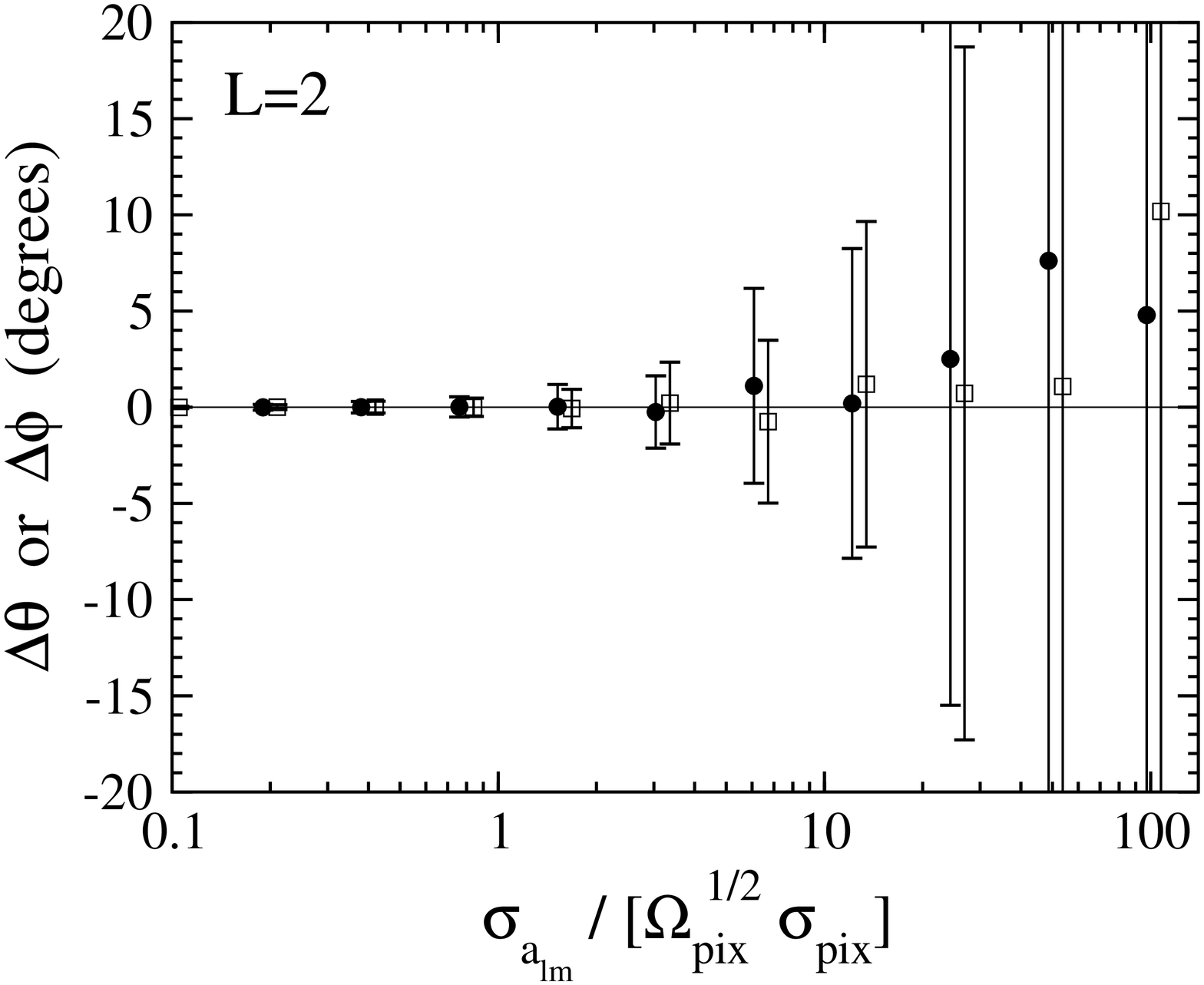}\\
\includegraphics[height=3.5in, width=2.5in, angle=-90]{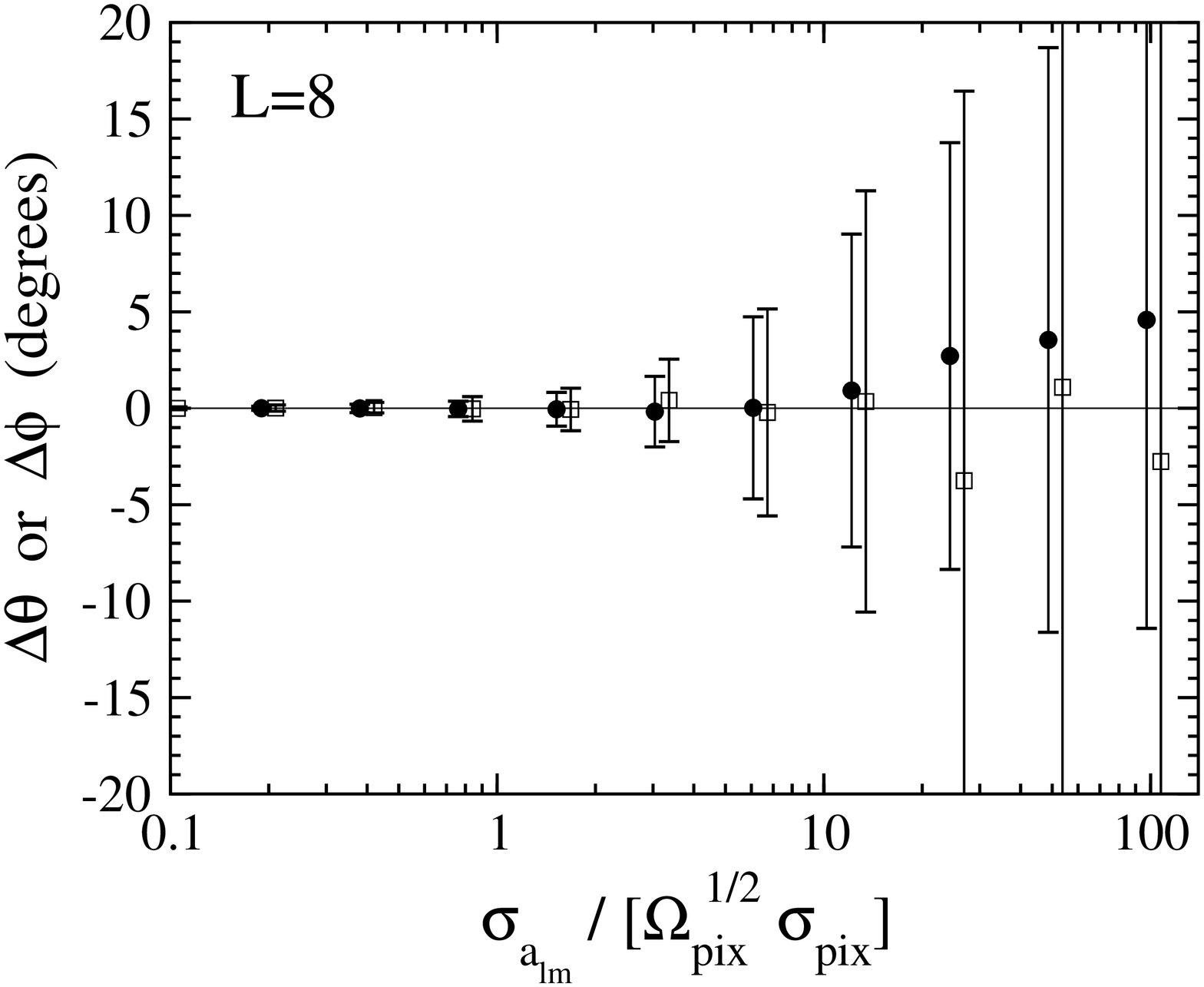}
\caption{The accuracy in $\theta$ (filled circles) and $\phi$ (empty
squares) of a chosen multipole vector as a function of noise added to
the $\alm$. Both the mean value and scatter in the shift of the angles
are shown. The $x$-axis value of unity corresponds to the scatter in
the $\alm$ due to WMAP V-band pixel noise.}
\label{fig:accuracy}
\end{figure}  

The mean and standard deviations of $\theta$ and $\phi$, as a function of
$\sigma_{\alm}$ (which is in units of WMAP V-band pixel noise
$\sqrt{\hat\Omega_{\rm pix}}\,\sigma_{\rm pix}$) are shown in
Fig.~\ref{fig:accuracy}. The two panels show the effects on the $\ell=2$
and $\ell=8$ vectors.  Note that both bias and scatter in $\theta$ and
$\phi$ can be read off this figure.  It is clear that the vectors are not
extremely sensitive to the accuracy in the $\alm$, and are determined to
about $\pm 1$ degree for noise which is of order pixel noise. If the noise
is much larger, however ($\sigma_{\alm}\gtrsim 10\sqrt{\hat\Omega_{\rm
    pix}}\,\sigma_{\rm pix}$), the accuracy in multipole vectors deteriorates
to the point that they are probably not useful as a representation of the
CMB anisotropy. Accurate determination of $\theta$ and $\phi$ is expected
to be especially important for higher multipoles, where the number of
vectors is large.

\section{Tests of Non-Gaussianity With Multipole Vectors}
\label{sec:tests}

Now that we have developed a formalism to compute the multipole
vectors, we would like to test the WMAP map for any unusual
features.  Clearly, there doesn't exist a test that simply checks for
the ``weirdness'' of any particular representation of the map, in this
case the multipole vectors. We can only test for features
that we specify in advance. Here our general goal is to test the
statistical isotropy of the map and search for any preferred directions.

Our motivation to devise and apply tests of statistical isotropy comes
from findings that the quadrupole and octupole moments of WMAP maps
lie in the plane of the Galaxy as discussed previously. One
could in principle extend this and ask whether any higher multipoles
lie preferentially in this (or any other) plane, and devise a
statistic to test for this alignment. Clearly, the number of tests one
can devise is very large.  Here we would like to be as general as
possible, and choose tests that suit our multipole vector
representation. We choose to consider the dot product of
multipole vectors between two different multipoles as described below.

\subsection{Vector Product Statistics }

A dot product of two unit vectors is a natural measure of their
closeness. One test we consider is dot products between all vectors
from the multipole $\ell_1$ with those from the multipole $\ell_2$. Since
our vectors are really ``sticks'' (i.e. each vector is determined only
up to a flip $\hat v\rightarrow -\hat v$), we always use the
absolute value of a dot product. Furthermore, motivated by the fact
that the WMAP quadrupole and octupole are located in the same plane and
that their axes of symmetry are only about 10$^\circ$ of each
other~\cite{angelica}, we also consider using the cross-products of
multipole vectors of any given multipole: if the vectors of $\ell_1$
and $\ell_2$ lie in a preferred plane, their respective cross
products are oriented near a common axis perpendicular to the plane. Dot
products of these two cross products would then be near unity.

We generalize this argument and make the following four choices for
our statistic, which we shall call $S$. For any two multipoles
$\ell_1$ and $\ell_2$, we consider the following.

\begin{enumerate}
  \item Dot products of multipole vectors $|\hat v^{\,(\ell_1, i)}\cdot
    \hat{v}^{\,(\ell_2, j)}|$, where $\ell_1\ne\ell_2$, $\hat{v}^{\,(\ell_1,
      i)}$ is the $i^{\rm th}$ vector from the $\ell_1$ multipole, and
    $\hat{v}^{\,(\ell_2, j)}$ is the $j^{\rm th}$ vector from the $\ell_2$
    multipole. We call this statistic ``vector-vector''. For a given $\ell_1$
    and $\ell_2$, there are clearly $M=\ell_1\ell_2$ distinct products.
    This statistic tests the orientation of vectors.
  \item Dot products $|\hat{v}^{\,(\ell_1, i)}\cdot (\hat{v}^{\,(\ell_2,
      j)}\times \hat{v}^{\,(\ell_2, k)})| / |\hat{v}^{\,(\ell_2,
      j)}\times \hat{v}^{\,(\ell_2, k)})|$ where where $\hat{v}^{\,(\ell_1,
      i)}$ comes from the $\ell_1$ multipole and $\hat{v}^{\,(\ell_2, j)}$
    and $\hat{v}^{\,(\ell_2,k)}$ from the $\ell_2$ multipole.  We call this
    statistic ``vector-cross''. For a given $\ell_1$ and $\ell_2$ and
    $j\neq k$, there are $M=\ell_1\ell_2(\ell_2-1)/2$ distinct products.
    This statistic tests the orientation of a vector with a plane.
  \item Dot products $|(\hat{v}^{\,(\ell_1, i)}\times \hat{v}^{\,(\ell_1,
      j)})\cdot (\hat{v}^{\,(\ell_2, k)}\times \hat{v}^{\,(\ell_2, m)})| /
    (|(\hat{v}^{\,(\ell_1, i)}\times
    \hat{v}^{\,(\ell_1,j)})|\,|(\hat{v}^{\,(\ell_2, k)}\times
    \hat{v}^{\,(\ell_2, m)})|)$
    where $\hat{v}^{\,(\ell_1, i)}$ and $\hat{v}^{\,(\ell_1, j)}$ come from
    the $\ell_1$ multipole and $\hat{v}^{\,(\ell_2, k)}$ and
    $\hat{v}^{\,(\ell_2, m)}$ come from the $\ell_2$ multipole.  We call
    this statistic ``cross-cross''. For a given $\ell_1$ and $\ell_2$
    and $i\neq j$ and and $k\neq m$, 
    there are $M=\ell_1(\ell_1-1)\ell_2(\ell_2-1)/4$ distinct products.
    This statistic tests the orientation of planes.
  \item Dot products $|(\hat{v}^{\,(\ell_1, i)}\times \hat{v}^{\,(\ell_1,
      j)})\cdot (\hat{v}^{\,(\ell_2, k)}\times \hat{v}^{\,(\ell_2, m)})|$
    where $\hat{v}^{\,(\ell_1, i)}$ and $\hat{v}^{\,(\ell_1, j)}$ come from
    the $\ell_1$ multipole and $\hat{v}^{\,(\ell_2, k)}$ and
    $\hat{v}^{\,(\ell_2, m)}$ come from the $\ell_2$ multipole.  We call
    this statistic ``oriented area''. For a given $\ell_1$ and $\ell_2$
    and $i\neq j$ and and $k\neq m$, 
    there are $M=\ell_1(\ell_1-1)\ell_2(\ell_2-1)/4$ distinct products.
    This statistic tests the orientation of areas.  Notice it is similar to
    the previous test but the cross products are unnormalized.
\end{enumerate}

\subsection{Rank-Order Statistic}\label{sec:rank-order}

\begin{figure*}[ht]
\includegraphics{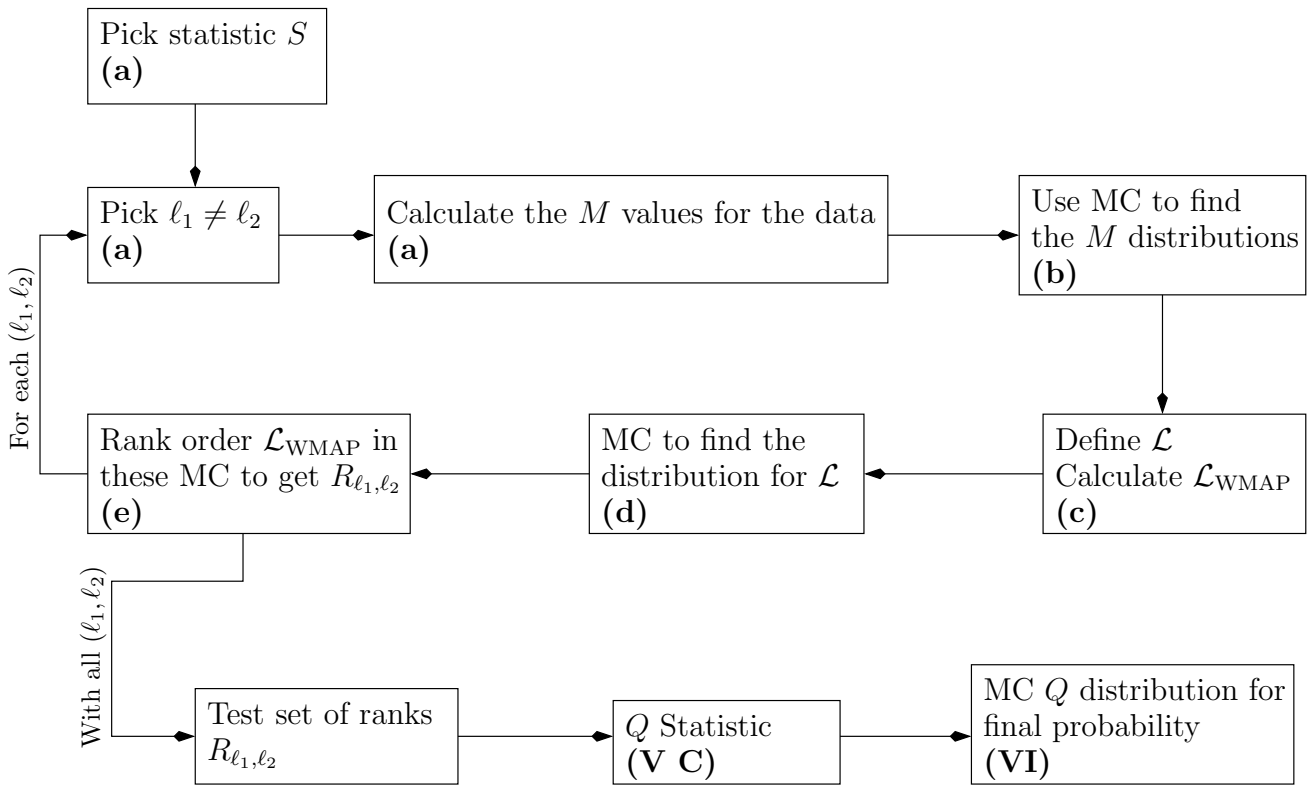}
\caption{A flowchart of the algorithm we apply in order to extract the
likelihood of the statistic, $S$.  The lower case bold faced letters, such as
{\bf (a)}, refer to the itemized points in section V B where more
information can be found about the step.  Also, the {\bf (V C)} and {\bf
  (VI)} refer to the sections in the paper where the details of these boxes
can be found.}
\label{fig:flowchart}
\end{figure*}  

Having computed the statistic in question, we would like to know the
likelihood of this statistic given the hypothesis that the $\alm$
are statistically isotropic. The most straightforward way, and
possibly the only reliable way, to do this is by comparing to
Monte Carlo (MC) realizations of the statistically isotropic {\it and\/}
Gaussian random $\alm$.  To be explicit, we provide the algorithm for
computing the rank-ordered likelihood, and also illustrate it in the flowchart
Fig.~\ref{fig:flowchart}.
\begin{description}
\item[(a)] For $\ell_1$ and $\ell_2$ fixed, a statistic $S$ will produce
  $M$ numbers (dot products).  We will use $S$ to test the hypothesis that
  the multipole vectors come from a map that exhibits statistical isotropy
  and Gaussianity.  Calculate the $M$ numbers for this statistic for the
  WMAP map.
\item[(b)] To determine the expected distributions for these $M$
  numbers we begin by generating 100,000 Monte Carlo Gaussian,
  isotropic maps; in other words, we draw the coefficients $\alm$ by
  assuming $\left\langle \alm a_{\ell'm'}^*\right\rangle = C_\ell
  \delta_{\ell\ell'} \delta_{m m'}$. We are assuming statistical
  isotropy {\it and\/} Gaussianity, and this is the hypothesis that we
  are testing.  We add a realization of inhomogeneous pixel noise,
  consistent with WMAP's V-band noise, to each MC map. For each MC
  realization, we then compute the multipole vectors for multipoles
  $\ell_1$ and $\ell_2$, and $M$ dot products of the statistic
  $S$. Because the vectors of any particular realization do not have
  an identity (e.g.\ we do not know which one is the ``fifth vector of
  $\ell=8$ multipole'') and neither do the dot products, we rank-order
  the $M$ dot products.  At the end we have $M$ histograms of the
  products, each having 100,000 elements.
\item[(c)] We would like to know the likelihood of the $M$ products
  computed from a WMAP map. To compute it we use a likelihood ratio test,
  which in the case of a single dot product of vectors would be the height
  of the histogram for the value of the statistic $S$ corresponding to
  WMAP relative to the maximum height. Since we have $M$ histograms, the
  likelihood trivially generalizes to
  \begin{equation}
    \mathcal{L}_{\rm WMAP}=\prod_{j=1}^{M} 
    \frac{N_{j, \rm WMAP}}{N_{j, \rm max}}
  \end{equation}
  where $N_{j, \rm WMAP}$ is the ordinate of the $j^{\rm th}$ histogram
  corresponding to WMAP's $j^{\rm th}$ rank-ordered product, $N_{j,\rm
    max}$ is the maximum value of $j^{\rm th}$ histogram, and the product
  runs over all $M$ histograms.
\item[(d)] Now that we have the WMAP likelihood, we need to compare it to
  ``typical'' likelihoods produced by Monte Carlo realizations of the map.
  (An alternative approach, computing the expected distribution of the
  likelihood from first principles, would be much more difficult since one
  would have to explicitly take into account the correlations between $M$
  products.) To do this, we generate another 50,000 Gaussian random
  realizations of coefficients $a_{\ell_1 m}$ and $a_{\ell_2 m}$, and
  compute the multipole vectors and the statistic $S$ for each realization.
\item[(e)] We rank-order the likelihood $\mathcal{L}_{\rm WMAP}$ among the
  50,000 likelihoods from MC maps to obtain its rank $R_{\ell_1, \ell_2}$.
\item[(f)] Finally we go to step (a) and repeat the whole procedure for all
  pairs of multipoles $(\ell_1, \ell_2)$ that we wish to test.  Only when
  we have the complete set do we assign a probability.
\end{description}

The rank $R_{\ell_1, \ell_2}$ gives the probability that the statistic
$S_{\rm WMAP}$ is consistent with the test hypothesis. For example, if
the likelihood of $S_{\rm WMAP}$ is rank $900$ out of $1000$, then there
is a 10\% probability that a Monte Carlo Gaussian random realization
of the CMB sky will give higher likelihood, and 90\% probability for a
lower MC likelihood. We say that the rank order of this particular
likelihood is 0.9. If our CMB sky is indeed random Gaussian, we would
expect the rank-orders of our statistics to be distributed between 0
and 1, being neither too small nor too large. Conversely, if we
computed the rank-orderings for three different multipoles for a
particular test and obtained 0.98, 0.99 and 0.95, (or 0.01, 0.05 and
0.02, say), we would suspect that this particular test is not
consistent with the Gaussian random hypothesis.

\subsection{How unusual are the ranks?}

We now consider how to quantify the confidence level for rejecting the
hypothesis of statistical isotropy of the $\alm$.  Let us assume that
we have computed the rank orders for $N$ different pairs of
multipoles, and obtained the rank-orderings of $x_i$ for the $i^{\rm
th}$ one, where $0\leq x_i\leq 1$. We consider the following
parametric test.

The test anticipates that the ranks will be unusually {\it high}. Let
us first order the ranks $x_i$ in descending order, so that $x_1$ is
the largest and $x_N$ the smallest.
We calculate the following statistic
\begin{equation}
Q(x_1, \ldots, x_N)=N! \int_{x_1}^1 dy_1\, \int_{x_2}^{y_1} dy_2 \ldots
        \int_{x_N}^{y_{N-1}} dy_N.
\label{eq:Q}
\end{equation}
If the ranks $R_{\ell_1, \ell_2}$ were expected to be uniformly
distributed in $[0, 1]$, $Q$ would be the probability that the highest
rank is greater than $x_1$, {\it and} the second-biggest rank greater
than $x_2$, \ldots, {\it and} the smallest rank greater than
$x_N$. However we do not expect that the ranks from Gaussian random
maps are uniformly distributed, and we treat $Q$ merely as a
statistic. We then ask that given the (possibly very small) value of
$Q_{\rm WMAP}$, what fraction of Gaussian random maps would give an
even smaller $Q$? That number is our probability, and is computed
in the next section. Note that, although apparently difficult or impossible to
evaluate analytically, the right-hand side of Eq.~(\ref{eq:Q}) can
trivially be computed using a recursion relation as shown in
appendix~\ref{app:prob}.

\section{Results}\label{sec:results}

Table I shows the final ranks for the vector-vector, vector-cross,
cross-cross, and oriented area tests for pairs ($\ell_1,
\ell_2$). These results correspond to the full-sky cleaned WMAP map
from Tegmark {\it et al.}~\cite{tegmark_cleaned} (results for their
Wiener-filtered map from the same reference are essentially
identical).  The ranks from the WMAP ILC full-sky map are similar, and
to make the presentation concise we show the ILC map oriented
area ranks in Table I, but otherwise only quote final probabilities
for the ILC ranks.  As discussed earlier, a detailed analysis of cut-sky
maps will be presented in an upcoming publication.

For computational convenience we considered only the multipoles
$2\leq \ell_1, \ell_2\leq 8$; we discuss the upper multipole limit in
section~\ref{sec:further_tests}.  Note that, for the vector-cross test,
($\ell_1, \ell_2$) and ($\ell_2, \ell_1$) products are distinct and
both need to be considered.

\begin{table*}[!t]
\begin{tabular}{||c||c|c||c|c|c||c|c||c|c||c|c||}\hline\hline
\multicolumn{12}{||c||}
        {\rule[-2mm]{0mm}{5mm} Ranks of Products of Multipole Vectors}\\\hline
           &              
\multicolumn{2}{|c||}{\rule[-2mm]{0mm}{6mm}Vector-Vector}  &     
\multicolumn{3}{|c||}{\rule[-2mm]{0mm}{6mm}Vector-Cross}& 
\multicolumn{2}{|c||}{\rule[-2mm]{0mm}{6mm}Cross-Cross} &
\multicolumn{2}{|c||}{\rule[-2mm]{0mm}{6mm}Oriented Area}&
\multicolumn{2}{|c||}{\rule[-2mm]{0mm}{6mm}O.\ A.\ (ILC map)}
\rule[-2mm]{0mm}{6mm}\\\hline
\rule[-2mm]{0mm}{6mm} ($\ell_1, \ell_2$) & 
$M$ & 
\hspace{0.3cm} Rank \hspace{0.3cm} &  
\hspace{0.cm}  $M$ \hspace{0.cm}  & 
\hspace{0.03cm} ($\ell_1, \ell_2$) Rank  \hspace{0.03cm} &
\hspace{0.03cm} ($\ell_2, \ell_1$) Rank  \hspace{0.03cm} & 
\hspace{0.02cm}  $M$ \hspace{0.02cm}  & 
\hspace{0.3cm}  Rank \hspace{0.3cm}  &
\hspace{0.02cm}  $M$ \hspace{0.02cm}  & 
\hspace{0.3cm}  Rank \hspace{0.3cm}  &
\hspace{0.02cm}  $M$ \hspace{0.02cm}  & 
\hspace{0.3cm}  Rank \hspace{0.3cm}  
\\ \hline
(2, 3) & 6    & 0.57714 & 6, 3     & 0.03176 & 0.04814 & 3   & 0.01316 & 3   & 0.00126 & 3   & 0.00886\\
(2, 4) & 8    & 0.39167 & 12, 4    & 0.51983 & 0.20369 & 6   & 0.56747 & 6   & 0.96042 & 6   & 0.62279\\
(2, 5) & 10   & 0.66656 & 20, 5    & 0.85252 & 0.10536 & 10  & 0.22285 & 10  & 0.71820 & 10  & 0.77484\\
(2, 6) & 12   & 0.53649 & 30, 6    & 0.50367 & 0.67791 & 15  & 0.87882 & 15  & 0.76320 & 15  & 0.93862\\
(2, 7) & 14   & 0.44925 & 42, 7    & 0.74254 & 0.52205 & 21  & 0.91890 & 21  & 0.86496 & 21  & 0.63763\\
(2, 8) & 16   & 0.21683 & 56, 8    & 0.20861 & 0.73486 & 28  & 0.91338 & 28  & 0.68520 & 28  & 0.99986\\
(3, 4) & 12   & 0.18093 & 18, 12   & 0.75272 & 0.30611 & 18  & 0.17059 & 18  & 0.34475 & 18  & 0.12562\\
(3, 5) & 15   & 0.21511 & 30, 15   & 0.36963 & 0.78578 & 30  & 0.37187 & 30  & 0.67870 & 30  & 0.76284\\
(3, 6) & 18   & 0.31507 & 45, 18   & 0.26683 & 0.54146 & 45  & 0.75052 & 45  & 0.93546 & 45  & 0.52839\\
(3, 7) & 21   & 0.98772 & 63, 21   & 0.85874 & 0.57072 & 63  & 0.55147 & 63  & 0.73650 & 63  & 0.83324\\
(3, 8) & 24   & 0.76120 & 84, 24   & 0.98578 & 0.60408 & 84  & 0.99988 & 84  & 0.99656 & 84  & 0.97766\\
(4, 5) & 20   & 0.41209 & 40, 30   & 0.28221 & 0.84716 & 60  & 0.54035 & 60  & 0.52936 & 60  & 0.65965\\
(4, 6) & 24   & 0.68840 & 60, 36   & 0.58372 & 0.86140 & 90  & 0.74826 & 90  & 0.62266 & 90  & 0.73762\\
(4, 7) & 28   & 0.85008 & 84, 42   & 0.51404 & 0.95584 & 126 & 0.53715 & 126 & 0.88992 & 126 & 0.32153\\
(4, 8) & 32   & 0.48723 & 112, 48  & 0.56328 & 0.85462 & 168 & 0.84374 & 168 & 0.99006 & 168 & 0.95266\\
(5, 6) & 30   & 0.82148 & 75, 60   & 0.42361 & 0.88662 & 150 & 0.66327 & 150 & 0.82760 & 150 & 0.86242\\
(5, 7) & 35   & 0.86884 & 105, 70  & 0.56542 & 0.96116 & 210 & 0.59483 & 210 & 0.68920 & 210 & 0.98440\\
(5, 8) & 40   & 0.83380 & 140, 80  & 0.96812 & 0.34287 & 280 & 0.30403 & 280 & 0.24449 & 280 & 0.33959\\
(6, 7) & 42   & 0.03742 & 126, 105 & 0.13831 & 0.02203 & 315 & 0.20221 & 315 & 0.97286 & 315 & 0.78414\\
(6, 8) & 48   & 0.92760 & 168, 120 & 0.91468 & 0.78058 & 420 & 0.72850 & 420 & 0.62894 & 420 & 0.66873\\\rule[-2mm]{0mm}{4.5mm}
(7, 8) & 56   & 0.03238 & 196, 168 & 0.04440 & 0.02060 & 588 & 0.09552 & 588 & 0.25367 & 588 & 0.24871\\
\hline\hline
\end{tabular}
\label{tab:ranks}
\caption{Ranks of the vector-vector, vector-cross, cross-cross, and
oriented area statistics for multipoles $2\leq \ell_1, \ell_2\leq 8$
for the four tests we consider as applied to the Tegmark {\it et
al.}~\cite{tegmark_cleaned} cleaned map. Also listed are the oriented
area ranks for the ILC map. $M$ is the number of products for a given
statistic for each $(\ell_1,\ell_2)$ pair. In
section~\ref{sec:oriented-area} we perform a parametric test to
compute the likelihood of oriented area ranks being this high.}
\end{table*}

\subsection{Vector-vector, vector-cross, and cross-cross ranks}

Table I shows that the vector-vector ranks are distributed roughly as
expected, nearly uniformly in the interval $[0, 1]$.  The vector-cross
ranks, however, are starting to shows hints of an interesting feature
that will be more pronounced later in the cross-cross and
oriented area tests: ranks that are unusually {\it high\/}: seven ranks
out of 42 are greater than 0.9. The probability of this happening,
however, is not statistically significant and may be purely
accidental.

The first big surprise comes from the cross-cross ranks: the
$(\ell_1=3, \ell_2=8)$ rank is $0.99988$.  This means that only 6 out
of 50,000 MC generated maps had a higher likelihood than the WMAP map!
In other words, the $84$ cross-cross dot products computed for this
pair from WMAP lie unusually near the peaks of their respective
histograms, which, recall, are built out of 100,000 products from MC
map realizations. {\it The violation of statistical isotropy and/or
Gaussianity therefore manifests itself by a particular correlation
between the vectors which makes the statistic $S_{\rm WMAP}$
``unusually usual''}. Since we checked that the distribution of
Monte-Carlo generated $(\ell_1=3, \ell_2=8)$ cross-cross ranks is
uniform in the interval $[0, 1]$, it is easy to see that the
probability of this rank being this high (or higher) is $6/50,000$, or
$0.012$\%.  Admittedly, we checked $21$ such cross-cross ranks, which
raises the probability of find such a result to $0.26$\%.   If we
include all vector-vector, vector-cross, cross-cross and oriented area
ranks, the probability rises to a still rather low $1.3$\%.

Note that this effect is very different from the now-familiar
orientation of the quadrupole and octupole axes; the
quadrupole-octupole alignment is quite {\it unlikely\/} and results in
multipole vector products which preferentially fall on the tails of
their respective histograms. This is confirmed by the actual ranks for
($2, 3$) and ($3, 2$) multipole vector-cross products, and also the
($2, 3$) cross-cross and oriented area products, all of which are
fairly low ($\lesssim 0.05$). This is easy to understand: in the
cross-cross product case, for example, the fact that the multipole
vectors lie mostly in the Galaxy plane implies that their cross
products are roughly perpendicular to this plane. The dot products of
those are then, by absolute value, very large, and hence unusual. What
we are seeing here is that the $(\ell_1=3, \ell_2=8)$ cross-cross
products from the WMAP full-sky map are {\it unusually usual}.

\subsection{Oriented area ranks}\label{sec:oriented-area}

\begin{figure}
\includegraphics[height=3.5in, width=2.8in, angle=-90]{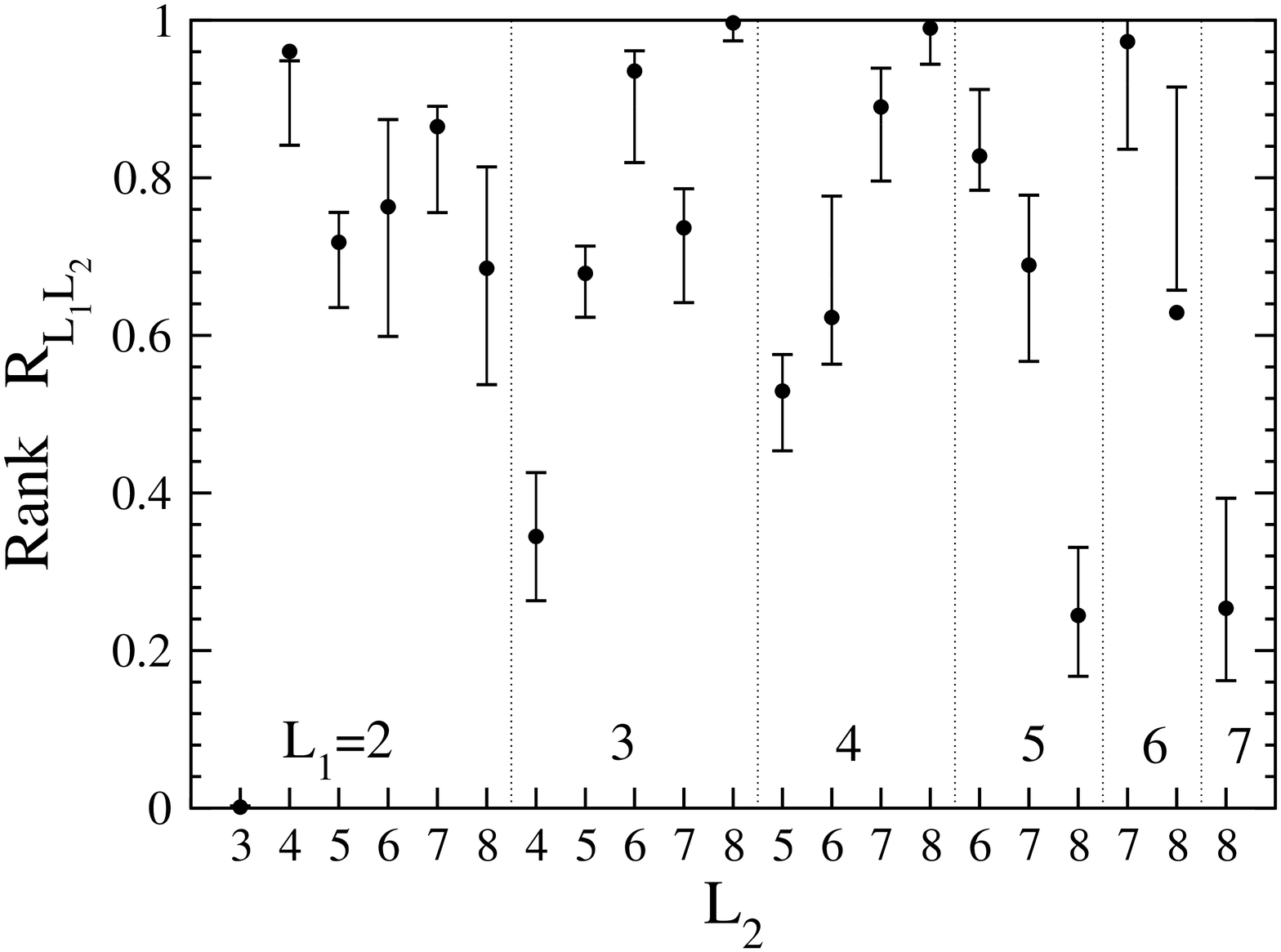}
\caption{Ranks of the oriented area statistics for the Tegmark full-sky
map. The mean values correspond to the actual extracted $\alm$,
while the error bars were obtained by adding the pixel noise (the
error bars are not necessarily symmetric around the no-error
values). Note that an unusually large fraction of the ranks are
high. Also note the {\it low\/} value of $\ell_1=2$, $\ell_2=3$ rank
which is due to the alignment of the quadrupole and octupole which
has been noted earlier~\cite{tegmark_cleaned}.}
\label{fig:cr_cr.ranks}
\end{figure}

As mentioned above, one of the cross-cross ranks was extremely high.
A much bigger surprise is found when we examine the
oriented area ranks (see also Fig.~\ref{fig:cr_cr.ranks}): 2 (out of 21)
are greater than 0.99, a total of 5 are greater than 0.9, and a total of 8
are greater than 0.8! These ranks are clearly not distributed in the same
way as those from a typical MC map.

To examine the probability of the ranks being this high we
perform the parametric test described in section~\ref{sec:rank-order}
and in appendix~\ref{app:prob}: we compute the statistic $Q$ which,
for a distribution of ranks expected to be uniform in $[0, 1]$,
would be the probability of the largest one being at least as large as the
largest actual WMAP rank {\it and} the second largest being at least
as large as the second-largest actual rank, {\it etc.} This statistic,
applied to ranks from the Tegmark {\it et al.}\ cleaned map, is
\begin{equation}
Q=10^{-(6.12^{+0.62}_{-1.16})}
\label{eq:wmap_q}
\end{equation}
where the error around the mean value is estimated by repeatedly
adding the pixel noise of the map to the pure extracted $\alm$, as
in Eq.~(\ref{eq:pixnoise}), and estimating the effect on the products
of multipole vectors and their ranks.

\begin{figure*}[ht]
\includegraphics[height=5.5in, width=4.5in, angle=-90]{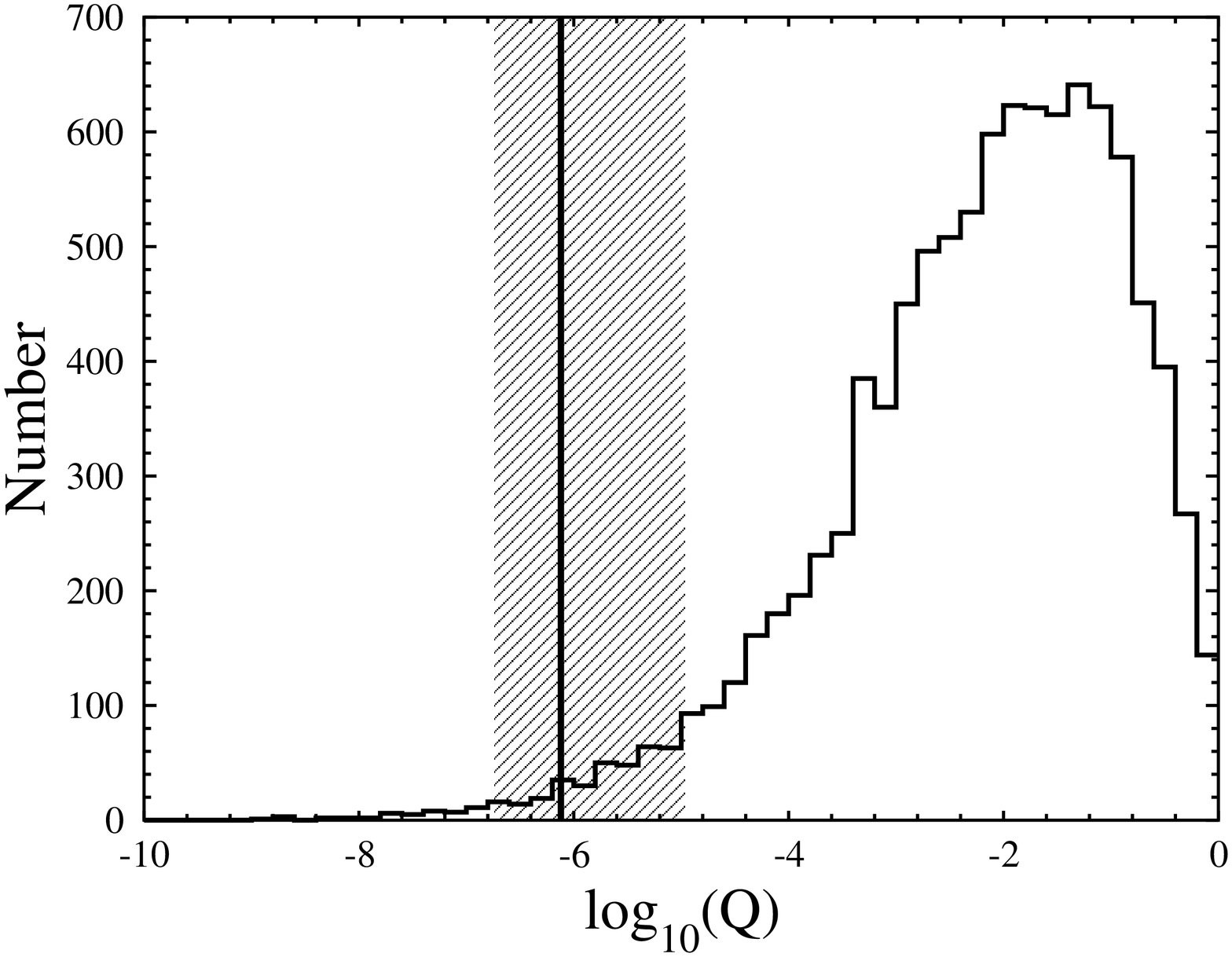}
\caption{The statistic $Q$ for the oriented area statistic, computed
from the Tegmark et al.\ cleaned WMAP map, is shown by vertical
line. The shaded region around it corresponds to the uncertainty due
to pixel noise, while the histogram shows the distribution of the
statistic for MC-generated Gaussian maps. Only $1.07\%$ of MC values
of $Q$ are smaller than the no-error value of the WMAP $Q$. The
same fraction for the ILC map (not shown here) is $0.38\%$.  }
\label{fig:hist_dotcross_final}
\end{figure*}

However, we have to be cautious not to over-interpret $Q$ as the final
probability---it has to be compared to Monte Carlo probabilities computed
under the same conditions to determine its distribution. To this end, we
generate 10,000 additional MC Gaussian random maps and compute $Q$ for
each. It turns out that only 107 of them produce $Q$ lower than the WMAP
value in~(\ref{eq:wmap_q}).  This is further illustrated in
Fig.~\ref{fig:hist_dotcross_final}, which also shows the error bars on the
WMAP $Q$. The probability of WMAP oriented area ranks being this high,
according to the $Q$-test, is 1.07\%, which corresponds to a 2.6-$\sigma$
(or $98.93\%$) evidence for the violation of statistical isotropy and/or
Gaussianity.  The connection of this (nearly) 3-$\sigma$ deviation
to the nearly 3-$\sigma$ deviation represented by the $(\ell_1=3, \ell_2=8)$
cross-cross rank remains somewhat unclear.

\subsection{Further tests}\label{sec:further_tests}

We have performed a few tests to explore the stability of the oriented
area result. First, we have varied the multipole range from the
fiducial $2\leq \ell\leq 8$; the results are shown in Table
II. Increasing the lower limit, $\ell_{\rm min}$, leads to the
final probability of $1\%$ and $5.6\%$ for $\ell_{\rm min}=3$ and 4
respectively. Therefore, evidence for the violation of statistical
isotropy weakens, but does so relatively slowly. This shows that our
main result does not completely hinge on the quadrupole and
octupole. Same can be said for the upper multipole limit, which gives
strongest results for the violation of statistical isotropy with
$\ell_{\rm max}=8$, but with decreasing $\ell_{\rm max}$ the result
does not immediately go away.  Finally, we have checked that
increasing $\ell_{\rm max}$ to higher values, up to 12, does not
produce new ranks that are unusually high.  The correlations are
therefore most apparent in the multipole range $2\leq (\ell_1,
\ell_2)\leq 8$.

\begin{table}
\begin{tabular}{||c|c|c||}\hline\hline
\multicolumn{3}{||c||}
        {\rule[-2mm]{0mm}{5mm} Varying the Multipole Coverage}\\\hline\hline
\rule[-2mm]{0mm}{6mm} 
$\ell_{\rm min}$ & \hspace{0.5cm} $Q_{\rm WMAP}$ \hspace{0.5cm}
& $f(Q_{\rm MC}<Q_{\rm WMAP})$ 
\\ \hline
\rule[-2mm]{0mm}{6mm} 2 & $7.61\times 10^{-7}$ & $107/10000$  \\
\rule[-2mm]{0mm}{6mm} 3 & $3.13\times 10^{-6}$ & $105/10000$ \\
\rule[-2mm]{0mm}{6mm} 4 & $3.12\times 10^{-4}$ & $565/10000$ \\
\hline\hline
\rule[-2mm]{0mm}{6mm} 
$\ell_{\rm max}$ & $Q_{\rm WMAP}$ & $f(Q_{\rm MC}<Q_{\rm WMAP})$ 
\\ \hline
\rule[-2mm]{0mm}{6mm} 8 & $7.61\times 10^{-7}$ &  $107/10000$ \\
\rule[-2mm]{0mm}{6mm} 7 & $3.72\times 10^{-5}$ &  $394/10000$ \\
\rule[-2mm]{0mm}{6mm} 6 & $3.62\times 10^{-3}$ &  $2079/10000$ \\
\hline\hline
\end{tabular}
\label{tab:l_range}
\caption{Final probabilities of the WMAP oriented area statistic as a
function of multipole coverage $l_{\rm min}\leq l\leq l_{\rm
max}$. The fiducial case is $2\leq \ell\leq 8$ and we have shown how
the results change if the lower and upper bounds are changed. $f(Q_{\rm
MC}<Q_{\rm WMAP})$ is the fraction of MC random Gaussian maps that
give a value of $Q$ smaller than WMAP.}
\end{table}

We next check if the correlations can be explained by any remaining
dust contamination.  We use WMAP's V-band map of the identified
thermal dust; this map was created by fitting to the template model
from Ref.~\cite{finkbeiner}. We assume for a moment that 10\% of the
identified contamination by dust had not been accounted for, and we
add it to the cleaned CMB map, i.e. $T_{\rm tot}=T_{\rm
CMB}+0.1\,T_{\rm dust}$.  Although we are adding a significant
contamination (the remaining dust is expected to contribute no more
than a few percent to the rms CMB temperature~\cite{wmap_results}),
the high oriented area ranks do not change much (they actually
slightly increase), and the Gaussian isotropic hypothesis is
still ruled out at the 99.3\% level. Clearly, dust contamination
does not explain our results.  Further, to mimic remaining foregrounds
due to an imperfect cleaning of the map we tested adding a synthetic
random Gaussian map which contributed 10\% of the RMS temperature. We
find that the oriented area statistic still disagrees with the
Gaussian isotropic hypothesis at the 99.4\% level.

Finally, we have repeated the analysis with several other available
full-sky CMB maps. As mentioned earlier, both maps analyzed by Tegmark
{\it et al.}\ give the same probability for the oriented area
statistic. For the WMAP ILC map we find similarly high ranks,
giving an even smaller value for our statistic: $Q_{\rm
ILC}=2.44\times 10^{-7}$, and only 62 MC maps out of 10000 have a
smaller value of $Q$; therefore, the high ranks in the ILC map are
unlikely at the 99.38\% level, corresponding to 2.7-$\sigma$.

\section{Discussion}

Does the apparent violation of statistical isotropy or Gaussianity
that we detected have a cosmological origin, or is it due to
foregrounds or measurement error?  The results presented in this paper
refer to full-sky maps, and it is known that there are two large cold
and two hot spots in the Galaxy plane, and that any result that
depends on structure in this plane is suspect. Nevertheless, it is far
from obvious that the result is caused by the contamination in the map
for the following reasons:

\begin{itemize}

\item As Tegmark {\it et al.}~\cite{tegmark_cleaned} argue, their cleaned
map agrees very well with the ILC map on large scales, although the
two were computed using different methods. The results we presented
indicate a violation of isotropy and/or Gaussianity at $\geq 98.9\%$ confidence 
using either map. Furthermore, instrumental noise and beam
uncertainties are completely subdominant on these scales.

\item The results come from an effect different than the quadrupole
and octupole alignment: the latter is fairly {\it unlikely}, as
discussed in Ref.~\cite{angelica}, while we see a particular
correlation between the vectors that makes our statistic $S$ unusually
{\it likely}.

\item Perhaps most importantly, our results are mostly (though not
completely) independent of the quadrupole and octupole, multipoles
that might be suspect. For example, the second-highest oriented area
rank is $(\ell_1=4, \ell_2=8)$. Furthermore, Table~\ref{tab:l_range}
shows that if we only use the multipoles $4\leq \ell_1, \ell_2\leq 8$,
the oriented area statistic still rules out the Gaussian random
hypothesis at the $94.4\%$ level.

\end{itemize}

At this time it is impossible to ascertain the origin of the
additional correlation between the multipole vectors that we are
seeing. One obvious way to find out more about their origin is to
Monte Carlo generate maps that are non-Gaussian or violate the
statistical isotropy, according to a chosen prescription, and see
whether our statistics, $S_{\rm WMAP}$, agree with the $S$ computed from MC
maps. Of course, there are many different ways in which Gaussianity
and/or isotropy can be broken, and there is no guarantee that we
can find one that explains our results.

Another possibility is to cut the galaxy (or other possible
contaminations) from the map prior to performing the vector
decomposition.  There are two approaches we can take: (1) use the
cut-sky $\alm$ to compute the multipole vectors and the statistics $S$
and compare those to $S$ computed from cut-sky Gaussian random maps,
or (2) reconstruct the true full-sky $\alm$ and compare with full-sky
Gaussian random maps. The latter procedure is preferred as one would
like to work with the true multipole vectors of our universe, but
reconstructing the full-sky map from the cut-sky information is a
subtle problem that will introduce an additional source of
error. Nevertheless, the total error with a $\sim 10^\circ$ cut may
still be small enough to allow using the vectors as a potent tool for
finding any preferred directions in the universe.  We are 
actively pursuing these approaches at the present time.

Finally, Park~\cite{park} recently tested the full-sky WMAP
maps using the genus statistic and found evidence for the violation of
Gaussianity at 2--3$\,\sigma$ level, depending on the smoothing scale and
the chosen aspect of the statistic.  Furthermore Eriksen {\it et
  al.}~\cite{NorthSouth} find that the WMAP multipoles with $\ell <35$ have
significantly less power in the northern hemisphere than in the southern
hemisphere.  It is possible that these results and the effects
discussed in this paper have the same underlying cause, but
this cannot be confirmed without further tests. 

\section{Conclusions}

The traditional $Y_{\ell m}$ expansion of the sky has many advantages.  For
one, each set of $Y_{\ell m}$ of fixed $\ell$ form an irreducible
representation of the rotation group in three dimensions $O(3)$; for
another, more than two centuries of effort have lead to a rich mathematical
literature on the $Y_{\ell m}$, their properties, and how to efficiently
calculate them.  The coefficients, $\alm$ of a $Y_{\ell m}$ expansion of a
function on the sphere are readily calculated as integrals over the sphere
of the function times the $Y_{\ell m}^*$.  In this paper we have considered
a different basis of an equivalent irreducible representations of the
proper rotation group, ${\cal O}_{(\ell)}^{i_1\cdots i_\ell}$ -- for each
$\ell$, the traceless symmetric product of $\ell$ copies of the unit vector
of coordinates ${\hat e}(\Omega)$.  These are merely linear combinations of
the $Y_{\ell m}$, and so share many of the properties of them, albeit with
a more sparse mathematical literature explicitly dedicated to their
properties.  In particular the coefficients $F^{(\ell)}_{i_1 \cdots
  i_\ell}$ of a ${\cal O}_{(\ell)}$ expansion of a function on the sky are,
like the spherical harmonic coefficients, calculable as integrals over the
sphere of the function times ${\cal O}_{(\ell)}$.

We have expressed the $F^{(\ell)}_{i_1 \cdots i_\ell}$ as symmetric
traceless products of $\ell$ (headless) unit vectors $\{\hat v^{(\ell,
  i)}\}$ and a scalar $A^{(\ell)}$.  The $\{\hat v^{(\ell, i)}\}$ are
highly non-linear functions of the $\alm$.  Thus, while in principle they
encode the exact same information, they may make certain features of the
data more self-evident.  In particular we claim that these ``multipole
vectors'' are natural sets of directions to associate with each multipole
of the sky.  A code to calculate multiple vectors from CMB skies is
available on our website at \website.

We have obtained the multipole vectors of the CMB sky as measured by
WMAP, as well as the oriented areas defined by all pairs of such
vectors (within a particular multipole).  We have examined the
hypothesis that the vectors of multipole $\ell$ are uncorrelated with
the vectors of multipole $\ell'$ for $\ell$ and $\ell'$ up to 8.  We
have done this by comparing in turn the dot products of the vectors
from $\ell$ with those from $\ell'$, the dots products of the vectors
with the unit normals to the planes.  the dot products of the unit
normals to the planes with each other, and the dot products of the
normals to the planes with each other, We have found that while there
is nothing unusual about the distribution of dot products of the
vectors with each other, the dot products of the normals to the planes
with each other (and, to a lesser extent, the dot products of the unit
normals to the planes with each other) are inconsistent with the
standard assumptions of statistical isotropy and Gaussianity of the
$\alm$.  To quantify this inconsistency we compared the distribution
of these dot products with those from $50,000$ Monte Carlo simulations
and found that they are inconsistent at the level of 107 parts in
10000 for the Tegmark {\it et al}.\ cleaned full-sky map and 62 parts
in 10000 for the ILC full-sky map.  These results are robust to the
inclusion of appropriate Poisson noise.  The sensitivity to a Galactic
cut will be explored in a future publication, but preliminary results
suggest that the results persist within the error bars, but eventually
decline in statistical significance as the uncertainties increase with
increasing cuts.

\begin{acknowledgments}
  We would like to thank Tom Crawford, Vanja Duki\'{c}, Doug
  Finkbeiner, Gary Hinshaw, Eric Hivon, Arthur Lue, Dominik Schwarz,
  David Spergel, Jean-Phillipe Uzan, Tanmay Vachaspati, and Ben
  Wandelt for helpful discussions. We also thank the anonymous referee
  for useful comments. We have benefited from using the publicly
  available Healpix package~\cite{healpix}. The work of the particle
  astrophysics theory group at CWRU is supported by the DOE.
\end{acknowledgments}

\appendix
\section{Vector decomposition equations}\label{app:vectordecomp}

The vector decomposition equations we derived~(\ref{eqn:vectordecomp})
can be recast in a numerically more convenient form.  The equations, as
written, involve complex valued coefficients.  Here we rewrite these
equations in terms of their purely real components.  To
begin we note that the spherical harmonics satisfy $Y_{\ell,-m} (\Omega) =
(-1)^m Y_{\ell m}^* (\Omega)$.  Thus the decomposition coefficients of a real
valued function, such as $\Delta T (\Omega)/T$, satisfy $a_{\ell,-m}^* =
(-1)^m \alm$.  This shows that all the information about the function is
encoded in the real part of $a_{\ell 0}$ (the imaginary part is identically
zero) and the real and imaginary parts of $\alm$ for $1\le m\le \ell$.
These are the $2\ell+1$ independent components we use in the vector
decomposition.  We thus only need to solve~(\ref{eqn:vectordecomp}) for
$0\le m\le \ell$.

For notational convenience we drop the $(\ell)$ superscript on
$a_{\ell-1,m-j}$, $b_{m'}$, and $\hat v$.  It should be understood that
these quantities are associated with a particular multipole and step in the
recursive decomposition procedure as outlined in
section~\ref{sec:vectordecomp}.  The correspondence between the dipole and
Cartesian coordinate directions~(\ref{eqn:dipoledir}) allows us to identify
$\hat v= (\hat v_{-1}, \hat v_0, \hat v_1)$ with standard coordinate axes via
\begin{eqnarray}
  \hat v_{-1} & = & \frac1{\sqrt2}\left( \hat v_x + i\hat v_y\right),
  \nonumber \\
  \hat v_0 & = & \hat v_z, \label{eqn:vecrelation}\\
  \hat v_{1} & = & -\frac1{\sqrt2}\left( \hat v_x - i\hat v_y\right). \nonumber
\end{eqnarray}
Finally the real and imaginary parts of the $\alm$ are
\begin{equation}
  \alm^{\rm re} = \frac12 \left( \alm + \alm^* \right)\quad \hbox{and}\quad
  \alm^{\rm im} = \frac1{2i} \left( \alm - \alm^* \right).
  \label{eqn:almparts}
\end{equation}
Applying~(\ref{eqn:vecrelation}) and~(\ref{eqn:almparts}) to the multipole
vector decomposition equations~(\ref{eqn:vectordecomp}) gives the following
equations
\begin{widetext}
  \begin{eqnarray}
    \alm^{\rm re} & = & C_0^{(l,m)} a_{\ell-1,m}^{\rm re} \hat v_z +
    \frac1{\sqrt2} C_{-1}^{(\ell,m)} \left( a_{\ell-1,m+1}^{\rm re} \hat v_x -
      a_{\ell-1,m+1}^{\rm im} \hat v_y \right)
    - \frac1{\sqrt2} C_{1}^{(\ell,m)} \left( a_{\ell-1,m-1}^{\rm re} \hat v_x +
      a_{\ell-1,m-1}^{\rm im} \hat v_y \right), \nonumber \\
    \alm^{\rm im} & = & C_0^{(l,m)} a_{\ell-1,m}^{\rm im} \hat v_z +
    \frac1{\sqrt2} C_{-1}^{(\ell,m)} \left( a_{\ell-1,m+1}^{\rm im} \hat v_x +
      a_{\ell-1,m+1}^{\rm re} \hat v_y \right)
    - \frac1{\sqrt2} C_{1}^{(\ell,m)} \left( a_{\ell-1,m-1}^{\rm im} \hat v_x -
      a_{\ell-1,m-1}^{\rm re} \hat v_y \right), \nonumber \\
    a_{\ell 0} = a_{\ell 0}^{\rm re} & = &
    C_0^{(l,0)} a_{\ell-1,0}^{\rm re} \hat v_z +
    \sqrt2 C_{1}^{(\ell,0)} \left( a_{\ell-1,1}^{\rm re} \hat v_x -
      a_{\ell-1,1}^{\rm im} \hat v_y \right), 
    \label{eqn:vecdecomp-solved}\\
    b_{m'}^{\rm re} & = & D_0^{(l,m)} a_{\ell-1,m}^{\rm re} \hat v_z +
    \frac1{\sqrt2} D_{-1}^{(\ell,m)} \left( a_{\ell-1,m+1}^{\rm re} \hat v_x -
      a_{\ell-1,m+1}^{\rm im} \hat v_y \right)
    - \frac1{\sqrt2} D_{1}^{(\ell,m)} \left( a_{\ell-1,m-1}^{\rm re} \hat v_x +
      a_{\ell-1,m-1}^{\rm im} \hat v_y \right), \nonumber \\
    b_{m'}^{\rm im} & = & D_0^{(l,m)} a_{\ell-1,m}^{\rm im} \hat v_z +
    \frac1{\sqrt2} D_{-1}^{(\ell,m)} \left( a_{\ell-1,m+1}^{\rm im} \hat v_x +
      a_{\ell-1,m+1}^{\rm re} \hat v_y \right)
    - \frac1{\sqrt2} D_{1}^{(\ell,m)} \left( a_{\ell-1,m-1}^{\rm im} \hat v_x -
      a_{\ell-1,m-1}^{\rm re} \hat v_y \right), \nonumber \\
    b_0 = b_0^{\rm re} & = &
    D_0^{(l,0)} a_{\ell-1,0}^{\rm re} \hat v_z +
    \sqrt2 D_{1}^{(\ell,0)} \left( a_{\ell-1,1}^{\rm re} \hat v_x -
      a_{\ell-1,1}^{\rm im} \hat v_y \right), \nonumber \\
    |\hat v| & = & \hat v_x^2 + \hat v_y^2 + \hat v_z^2 = 1. \nonumber
  \end{eqnarray}
\end{widetext}
  Note that the equations for the $b_{m'}$ are identical to those for
  $\alm$ with $D^{(\ell,m)}$ inserted in place of $C^{(\ell,m)}$.  Here
  $1\le m\le \ell$ and $1\le m'\le \ell-2$.  These equations involve only
  real quantities and can thus be easily coded and solved.  These are the
  equations we have implemented to find the multipole vectors.

\section{Probability of rank orderings}\label{app:prob}

Consider $N$ numbers $x_i$, where $0\leq x_i\leq 1$, and order them in
descending order, so that $x_1$ is the largest and $x_N$ the smallest. Let
us then consider a set of variates $y_i$ uniformly distributed in the
interval $[0, 1]$, and also order them in descending order so that $y_1$ is
the largest one and $y_N$ the smallest. We ask: what is the probability
that $y_1$ is greater than $x_1$, {\it and\/} that $y_2$ is greater than
$x_2$, \ldots, {\it and\/} the $y_N$ is greater than $x_N$.

The probability that $y_1$ is in the interval $[x_1,
x_1+dx_1]$, $\mathcal{P}_1(x_1) dx_1$, is
\begin{equation}
\mathcal{P}_1(x_1) dx_1={N\choose 1}x_1^{N-1}dx_1
\end{equation}
and the probability that $y_1$ is larger than $x_1$, $P_1(x_1)$, is obviously
\begin{equation}
P_1(x_1)=\int_{x_1}^1 \mathcal{P}_1(y_1) dy_1.
\end{equation}

Given that $y_1$ is greater than $x_1$, the probability that the
$y_2$ is in the interval $[x_2, x_2+dx_2]$
\begin{equation}
\mathcal{P}_2(x_2|x_1) dx_2={N-1\choose 1}\,\left (\frac{x_2}{x_1}\right )^{N-2}
                                                \,\frac{dx_2}{x_1}
\end{equation}
and the probability that the largest $y$ is greater than $x_1$ {\it
and\/} the second-largest greater than $x_2$ is 
\begin{eqnarray}
P_2(x_1, x_2) &=& \int_{x_1}^1     \mathcal{P}_1(y_1)     dy_1     
                  \int_{x_2}^{y_1} \mathcal{P}_2(y_2|y_1) dy_2 \nonumber \\
        &=& N(N-1)  \int_{x_1}^1  dy_1  \int_{x_2}^{y_1} y_2^{N-2} dy_2.
\end{eqnarray}
We can continue this argument for all other $y_i$ and $x_i$, in descending
order in $x_i$. The final probability, the  joint  probability
of $i^{\rm th}$ largest $y$ being greater than $x_i$ for all $i$, is given by
\begin{equation}
P_N(x_1, x_2, \ldots, x_N) =  N!
        \int_{x_1}^1 dy_1  \int_{x_2}^{y_1}dy_2 \ldots 
        \int_{x_{N-1}}^{y_{N-1}} dy_N.
\label{eq:prob_parametric}
\end{equation}

We would like to evaluate this integral. Even though the result will
obviously be a polynomial in $x_i$,  there is a total of $2^N$ terms and
it is difficult to do the bookkeeping. However, there is a simple recursion
formula for this integral. Assume, more generally, that we want to
compute
\begin{equation}
I_N^{\alpha} \equiv 
        \int_{x_1}^1 dy_1  \int_{x_2}^{y_1}dy_2 \ldots 
        \int_{x_{N-1}}^{y_N} dy_N\, y_N^{\alpha}.
\end{equation}
One can then perform the innermost integral, and this leads to the
recursion relation
\begin{equation}
I_N^{\alpha}= \frac1{\alpha+1}\, \left[
  I_{N-1}^{\alpha+1}-x_N^{\alpha+1}I_{N-1}^0 \right].
\label{eq:recursion}
\end{equation}

We are left with two $(N-1)$-tuple integrals.
Therefore, starting from the $N$-dimensional integral, we can
recursively bring it down all the way to $N=1$, at which point it is
an easy one-dimensional integral
\begin{equation}
I_1^{\beta}\equiv \int_{x_1}^1 dy_1 \,y_1^{\beta}=
        \frac1{\beta+1} \left[ 1-x_1^{\beta+1} \right].
\label{eq:rec_last}
\end{equation}
for the required $\beta$. Using the recursion relation
(\ref{eq:recursion}), together with (\ref{eq:rec_last}), we numerically
compute the probability in~(\ref{eq:prob_parametric}).


\begin{thebibliography}{99}



\bibitem{wmap_results}
       C.L. Bennett {\it et.al.}, Astrophys. J. {\bf 148}, S1 (2003)

\bibitem{wmap_foreground}
       C.L. Bennett {\it et.al.}, Astrophys. J. {\bf 148}, S97 (2003)

\bibitem{wmap_angps}
       G. Hinshaw {\it et.al.},  Astrophys. J. {\bf 148}, S135 (2003)

\bibitem{wmap_low}
       D.N. Spergel {\it et al.}, Astrophys. J. {\bf 148}, S175 (2003)

\bibitem{maxima}
        S. Hanany {\it et al.}, Astrophys. J. {\bf 545}, L5 (2000)
  
\bibitem{DASI}
        N.W. Halverson {\it et al.}, Astrophys. J. {\bf 568}, 38 (2002)
 
\bibitem{acbar}
        C.L. Kuo {\it et al.}, Astrophys. J. {\bf 600}, 32 (2004) 
 
\bibitem{CBI}
        T.J. Pearson {\it et al.}, Astrophys. J. {\bf 591}, 556 (2003)
                                                                                
\bibitem{vsa}
        A.C. Taylor {\it et al.}, Mon. Not. R. Astron. Soc. {\bf 341}, 1066
        (2003)
 
\bibitem{archeops}
        A. Benoit {\it et al.}, Astron. Astrophys. {\bf 399}, L19 (2003)
 
\bibitem{boom}
        J.E. Ruhl {\it et al.}, Astrophys. J. {\bf 599}, 786 (2003)

\bibitem{scott}
        E.F. Bunn and D. Scott, Mon. Not. R. Astron. Soc. {\bf 331}, 313 (2000)

\bibitem{tegmark_cleaned}
        M. Tegmark, A. de Oliveira-Costa and A.J.S. Hamilton, 
        Phys. Rev. D,  {\bf 68}, 123523 (2003)

\bibitem{NorthSouth}
        H.K. Eriksen, F.K. Hansen, A.J. Banday, K.M. Gorski,
        and P.B. Lilje., submitted (astro-ph/0307507)

\bibitem{park}
        C.-G. Park, Mon. Not. R. Astron. Soc., {\bf 349}, 313 (2004)

\bibitem{Hajianetal}
       A. Hajian, T. Souradeep, N. Cornish, D.N. Spergel, and G.D. Starkman,
       in preparation

\bibitem{HajianSouradeep}
       A. Hajian and T. Souradeep, submitted to PRL (astro-ph/0301590)

\bibitem{vielva}
       P. Vielva, E. Martinez-Gonzalez, R.B. Barreiro, J.L. Sanz and
       L. Cayon, astro-ph/0310273

\bibitem{defects}
        F.R. Bouchet, D.P. Bennett and A. Stebbins, Nature
        335, {\bf 410} (1988)

\bibitem{salopek_bond}
       D.S.\ Salopek and J.R.\ Bond,  Phys.\ Rev.\ D {\bf 43}, 1005 (1991)

\bibitem{falk}
       T.\ Falk, R.\ Rangarajan and M.\ Srednicki,
       Astrophys.\ J.\ {\bf 403}, L1 (1993)

\bibitem{gangui}
       A. Gangui, F. Lucchin, S. Matarrese and S. Mollerach, 
       Astrophys.\ J.\ {\bf 430}, 447 (1994)

\bibitem{wang_kam}
        L. Wang and M. Kamionkowski, Phys.\ Rev.\ D {\bf 61}, 063504 (2000)

\bibitem{bern_uzan}
       F.\ Bernardeau and J.-P.\ Uzan, Phys.\ Rev. D {\bf 66}, 103506 (2002)

\bibitem{luo_schramm}
       X.\ Luo and D.N.\ Schramm, Phys.\ Rev.\ Lett.\ {\bf 71}, 1124 (1993)

\bibitem{luo}
       X.\ Luo,  Astrophys.\ J.\ {\bf 427}, L71 (1994)

\bibitem{munshi}
       D. Munshi, T.\ Souradeep and A.A.\ Starobinsky,  
       Astrophys.\ J.\ {\bf 454}, 552 (1995)

\bibitem{gold_spergel}
       D.M. Goldberg and D.N. Spergel, Phys. Rev. D {\bf 59}, 103002 (1999)

\bibitem{cooray_hu}
       A.\ Cooray and W.\ Hu, Astrophys.\ J.\ {\bf 534}, 533 (2000)

\bibitem{cooray_sz}
       A.\ Cooray, Phys.\ Rev.\ D.\ {\bf 64}, 063514 (2001)

\bibitem{heavens}
       A.F.\ Heavens, Mon. Not. R. Astron. Soc. {\bf 299}, 805 (1998)

\bibitem{spergel_gold}
       D.N. Spergel and D.M. Goldberg, Phys. Rev. D {\bf 59}, 103001 (1999)

\bibitem{verde}
       L. Verde, L.\ Wang, A.F.\ Heavens and M.\ Kamionkowski, 
       Mon. Not. R. Astron. Soc. {\bf 313}, 141 (2000)

\bibitem{komatsu_cobe}
       E.\ Komatsu {\it et al.}, Astrophys.\ J.\ {\bf 566}, 19 (2002)

\bibitem{santos_prl}
        M.G.\ Santos {\it et al.}, Phys.\ Rev.\ Lett.\ {\bf 88}, 241302 (2002)

\bibitem{santos_big}
        M.G.\ Santos {\it et al.}, Mon. Not. R. Astron. Soc. {\bf 341}, 623 (2003)

\bibitem{komatsu_wmap}
        E.\ Komatsu {\it et al.}, Astrophys.\ J.\ {\bf 148}, 119S (2003)

\bibitem{hu_trispec}
       W.\ Hu,  Phys.\ Rev.\ D.\ {\bf 64},  083005 (2001)

\bibitem{detroia}
       G. De Troia {\it et al.}, Mon. Not. R. Astron. Soc. {\bf 343}, 284 (2003)

\bibitem{gott}
        J.R.\ Gott {\it et al.}, Astrophys.\ J.\ {\bf 352}, 1 (1990)

\bibitem{smoot_genus}
       G. Smoot {\it et al.}, Astrophys. J. {\bf 437}, 1 (1994)

\bibitem{winitzki}
       S. Winitzki and A.\ Kosowsky, New Astronomy {\bf 3}, 75 (1997)

\bibitem{dolgov}
       A. Dolgov, A. Doroshkevich, D. Novikov and I. Novikov, 
       Int.\ J. Mod. Phys. D {\bf 8}, 189 (1999)
       

\bibitem{park_teg}
       C.-G.\ Park, C.\ Park, B.\ Ratra and M.\ Tegmark,   
       Astrophys.\ J.\ {\bf 556}, 582 (2001)

\bibitem{barreiro}
       R.B.\ Barreiro {\it et al.}, Mon. Not. R. Astron. Soc. {\bf 318}, 475 (2000)

\bibitem{mukherjee}
       P.\ Mukherjee, M.P.\ Hobson and A.N.\ Lasenby, 
       Mon. Not. R. Astron. Soc. {\bf 318}, 1157 (2000)

\bibitem{kogut}
       G. Kogut {\it et al.}, Astrophys.\ J.\ {\bf 464}, L29 (1996)

\bibitem{bromley}
       B.C. Bromley and M. Tegmark, Astrophys. J. {\bf 524}, L79 (2000)

\bibitem{wu}
        J.H.P.\ Wu {\it et al.}, Phys.\ Rev.\ Lett.\ {\bf 87}, 251303 (2001)

\bibitem{savage}
       R.\ Savage {\it et al.}, Mon. Not. R. Astron. Soc., submitted
       (astro-ph/0308266)

\bibitem{polenta}
       G. Polenta {\it et al.},   Astrophys.\ J.\ {\bf 572}, L27 (2002)

\bibitem{hansen}
       F. Hansen, D. Marinucci, P. Natoli and N. Vittorio, Phys.\ Rev.\ D
       {\bf 66}, 063006 (2002)

\bibitem{dore}
       O. Dor\'{e}, S.\ Colombi and F.R. Bouchet,
       Mon. Not. R. Astron. Soc. {\bf 344}, 905 (2003)

\bibitem{chiang}
       L.-Y. Chiang, P. Naselsky, O.V.\ Verkhodanov and M.\ Way, 
       Astrophys.\ J.\ {\bf 590}, L65 (2003)

\bibitem{coles}
        P. Coles, P. Dineen, J. Earl and D. Wright,
        MNRAS submitted, astro-ph/0310252

\bibitem{ferreira}
        P.\ Ferreira, J.\ Magueijo and K.\ Gorski, 
        Astrophys.\ J.\ {\bf 503}, L1 (1998)

\bibitem{zaroubi}
        A.J. Banday, S. Zaroubi and K.M. Gorski, Astrophys. J. {\bf 533},
        575 (2000).

\bibitem{magueijo}
        J. Magueijo, Astrophys.\ J.\ {\bf 528}, L57 (2000)

\bibitem{CSS}
       N.J. Cornish, D.N. Spergel and G.D. Starkman, 
       Class. Quant. Grav. {\bf 15}, 2657 (1998)

\bibitem{DMR1}
       G.F. Smoot, {\it et al.}, Astrophys. J. {\bf 396}, L1 (1992).

\bibitem{DMR4}
       G. Hinshaw {\it et al.}, Astrophys. J. {\bf 464}, L25 (1996)

\bibitem{efstathiou}
       G. Efstathiou, Mon. Not. R. Astron. Soc. {\bf 348}, 885 (2004)

\bibitem{angelica}
        A. de Oliveira-Costa, M. Tegmark, M. Zaldarriaga and A. Hamilton,
        astro-ph/0307282

\bibitem{wmap_suppl}
       M. Limon {\it et.al.}, ``WMAP Explanatory Supplement'', 
       http://lambda.gsfc.nasa.gov/product/map/m\_docs.cfm

\bibitem{healpix}
        K. Gorski, E. Hivon and B.D. Wandelt, Proceedings of the MPA/ESO 
        Cosmology Conference "Evolution of Large-Scale Structure", 
        eds. A.J. Banday, R.S. Sheth and L. Da Costa, 
        PrintPartners Ipskamp, NL, pp. 37-42 (astro-ph/9812350)

\bibitem{finkbeiner}
       D.P. Finkbeiner, M. Davis and D.J. Schlegel,
       Astrophys. J. {\bf 524}, 867 (1999)

\end{thebibliography}
\end{document}